%% file: HolSFM-Formal.tex
\documentclass[a4paper,12pt]{scrartcl}
\usepackage[utf8x]{inputenc}
\usepackage{color}

\include{Macros}

\usepackage{graphicx}

\usepackage{units}

\newcommand{\coh}{\nicefrac{1}{2}}
\renewcommand{\H}{H} 

\newcommand{\bline}{\vskip 1cm} 

\title{Holonomy Spin Foam Models: Boundary Hilbert Spaces and Time Evolution Operators}
\author{Bianca Dittrich$^{1,2}$, Frank Hellmann$^2$, Wojciech Kami{\'n}ski$^2$\\
\small $^1$ Perimeter Institute, 
\small  31 Caroline St. N, Waterloo, ON N2L 2Y5, Canada\\
\small   $^2$ MPI f. Gravitational Physics, 
 \small Am M\"uhlenberg 1, D-14476 Potsdam, Germany
 }

 \date{}

\begin{document}

\maketitle

\begin{abstract}
In this and the companion paper a novel holonomy formulation of so called Spin Foam models of lattice gauge gravity are explored. After giving a natural basis for the space of simplicity constraints we define a universal boundary Hilbert space, on which the imposition of different forms of the simplicity constraints can be studied. We detail under which conditions this Hilbert space can be mapped to a Hilbert space of projected spin networks or an ordinary spin network space.

These considerations allow to derive the general form of the transfer operators which generates discrete time evolution. We will describe the transfer operators for some current models on the different boundary Hilbert spaces and highlight the role of the simplicity constraints determining the concrete form of the time evolution operators.

\end{abstract}

\section{Introduction}

Together with the companion paper \cite{Comp} this paper introduces and studies a class of lattice
gauge theories that ocurr as spin foam models of quantum gravity
\cite{Reisenberger:1994aw,Reisenberger:1997sk,Baez:1997zt,Barrett:1997gw,Baez:1999sr,Barrett:1999qw,
Oeckl:2001wm,Oeckl:2005rh,Engle:2007uq,Engle:2007wy,Engle:2007qf,Freidel:2007py,Bahr:2010bs,
Perez:2012wv,Bianchi:2012nk}.

As opposed to previous formulations that stressed the relationship to the loop quantum gravity Hilbert space \cite{Rovelli:1989za,Ashtekar:1991kc,Ashtekar:1994mh,Ashtekar:1994wa,Ashtekar:1995zh,Rovelli:1995ac,Lewandowski:2005jk}, and thus was given in terms of spin networks and their geometric interpretation, we will here focus on a presentation that is as close as possible to lattice gauge theory. There is some overlap with the formulations explored in the context of auxilliary field theories on the group manifold, \cite{Boulatov:1992vp,Ooguri:1992eb,Reisenberger:1997sk,Reisenberger:1996pu,DePietri:1999bx,Reisenberger:2000fy,Reisenberger:2000zc,Oriti:2006se,Geloun:2010vj,Gurau:2009tw,Baratin:2011hp}, though our perspective is markedly different. Further there are similarities to the formulations of spin foam models in connection variables \cite{Oriti:2000hh,Perez:2000fs,Perez:2000ec,Pfeiffer:2001ny,Oriti:2002hv,Livine:2002rh,Barrett:2011qe} and as more ordinary discretized theories \cite{Freidel:1998pt,Oeckl:2005rh,Freidel:2007py,Bonzom:2009wm,Bonzom:2009hw}, and non-commutative first order formulations \cite{Baratin:2011tx,Baratin:2011hp}. In a forthcoming work \cite{AD} we will show how to extract geometric meaning directly from the formulation given here.

The formulation we explore here takes as its starting point the heuristic formulation of lattice BF theory on a 2-complex. We then insert simplicity constraints in the holonomy picture of this formulation, assuming the most general form common to most simplicity constraints. This allows us to define a very natural space of simplicity constraints covering almost all models in the literature, the exception being the model due to KKL \cite{Kaminski:2009fm,Kaminski:2009cc}. Combined with the results on the structure of the simplicity constraints of spin foam models in \cite{AD} this enables us to give natural extensions of the EPRL/FK and BC model to arbitrary, including finite, groups that we explored in \cite{Comp}.

The key aim of this paper is to explore the consequences of this generic formalism. We will give an explicit basis for the space of simplicity constraints in section \ref{Sec:BasisSpaceOfTheories}, and discuss which models lead to real partition functions. In section \ref{sec-BoundAndHilb} we then explore different ways of introducing boundaries and composing 2-complexes. These will lead to different notions of boundary Hilbert spaces. Next to the known spin network \cite{Rovelli:1995ac} and projected spin network \cite{Alexandrov:2002xc,Livine:2002ak} spaces we introduce a new universal boundary space common to all spin foam models built on the same group. We will also show how the assumptions we make on the structure of the simplicity constraints translates to the operator spin foam formalism, show how the usual models fall into this category and give the basis coefficients of the established models in the space of simplicity constraints in section \ref{sec-StandardSFM}.

In section \ref{transfer} we use the results obtained so far, to derive the general form of the transfer operator for any holonomy spin foam model in the different boundary Hilbert spaces. Transfer operators arise from a space--time decomposition of the partition function and indeed generate the (discrete) time evolution on the given Hilbert space. Thus transfer operators can be seen as the discrete time equivalent of Hamiltonian operators. Indeed for standard lattice systems Hamiltonians can be derived from transfer operators by a limiting procedure. We will shortly touch on the issue of how to take the limit in section \ref{transfer}. The general form of the transfer operator allows to illuminate the dynamics of spin foams, to highlight the role of the simplicity constraints and to clarify the connections between the different boundary Hilbert spaces.

We end with a discussion and outlook in section \ref{discussion}.

\section{Holonomy Formulation of Spin Foam Models}\label{sec-HolForm}

We will begin by recalling the formulation of spin foam models based on two arbitrary finite or Lie
groups $G$ and $\H \subset G$ in terms of holonomies on an arbitrary 2-complex.

\subsection{The Data}

We start with an arbitrary, finite, combinatorial 2-complex $\CC$. The 2-complex $\CC$ consists of faces, edges and vertices and we write $f \in \CC_f$, $e \in \CC_e$ and $v \in \CC_v$ respectively.

Each face of the 2-complex comes with a fiducial orientation given by the order of edges and vertices around it, as well as a fiducial base vertex. This orientation is unrelated to any orientation on a manifold from which the 2-complex might be constructed. The combinatorial information, together with the fiducial orientation and base vertex can be conveniently encoded by thinking of a face as an ordered set of the vertices and edges bordering it, $f = (v,e,v',e',\dots\, v)$. The notation $(a,b) \subset f$ will always mean that the ordered set $(a,b)$ exists as an uninterrupted subset in the ordered set $f$, that is $f = (v, \dots ,a,b, \dots v)$. We will similarly write $v \in e$ and $e \in f$ to denote adjecancy relationships.

As data on these 2-complexes we introduce one $G$ element $g_{ev} = g_{ve}^{-1}$ per half edge, and one $G$ element $g_{ef}$ per neighbouring edge and face, $e\in f$ \footnote{Note that these $g_{ef}$ have nothing to do with the holonomy from the mid point of the edge to the midpoint of the face that is introduced in the wedge formalism.}. In order to have a natural composition of group elements this should be read from right to left, that is, the group elments $g_{ab}$ and $g_{bc}$ associated to the ordered sets $(a,b)$ and $(b,c)$ respectively compose naturally to $g_{ab}g_{bc}$ if we read $g_{ab}$ as going from $b$ to $a$ and $g_{bc}$ as going from $c$ to $b$.

From these data we define two types of face holonomies as,

\bea\label{eq-FaceHolos}
g_f &=& \prod_{\substack{(a,b,c)\subset f\\b \in \CC_e}} g_{ab} g_{bf} g_{bc}\nn\\
\tilde{g}_f &=& \prod_{(a,b) \subset f} g_{ab},
\eea

Or, more eliptically, $g_f = g_{ve} g_{ef} g_{ev'} g_{v'e'} g_{e'f} \dots$ and $\tilde{g}_f = g_{ve} g_{ev'} g_{v'e'} \dots$

\subsection{The Partition Function}

We will first consider the partition function for a 2-complexes without boundary, that is, we treat all edges and faces as internal. We will see later that most spin foam models considered in the literature fall under the following definition:


%
\begin{defi}[Holonomy Spin Foam Model (no boundary)]\label{def-HolSFMNoBo}

Let $\CC$ be a 2-complex with orientations and base points on the faces, $G$ a unimodular Lie or
discrete group with Haar meassure $\dd g$, and $\H \subset G$ a subgroup of $G$. Then given two
complex valued distributions $E$ and $\omega$ on $G$ satisfying
\begin{itemize}
 \item $E(h g (h)^{-1}) = E(g)$ for all $h \in \H$,
 \item $E(g) = E(g^{-1})$ and $\omega(g) = \omega(g^{-1})$,
 \item $\omega(g) = \omega(\tilde{g} g \tilde{g}^{-1})$ for all $\tilde{g} \in G$,
\end{itemize}

we define the Holonomy Spin Foam partition function defined by $E$ and $\omega$ to be:

\be\label{eq-PartFuncNoBo}
\ZZ(\CC) = \int \left(\prod_{e \subset f}  \dd g_{ef}\right) \left( \prod_{v \subset e} \dd g_{ev} \right) \left(\prod_{e \subset f} E(g_{ef})\right) \left(\prod_f \omega(g_f)\right).
\ee

If we further have $\omega(g) = \overline{\omega(g)}$ and $E(g) = \overline{E(g)}$ we say that we have a Real Holonomy Spin Foam Model.

\end{defi}

An immediate consequence of the properties of $E$ and $\omega$ is that $\ZZ(\CC)$ is independent of the orientations and base vertices chosen.

Note that the partition function as given will usually diverge, even for compact groups. For noncompact groups there is also an ambiguity in the normalisation of the Haar measure and additional divergences due to gauge orbits. However, if $\omega$ is a regular function and the group is compact the model will be well defined. For finite groups this is always the case.

In the rest of this section we will study some general consequences of this definition, arising mostly from the structure of the integrand. Thus we will ignore issues of convergence from here on.

\subsection{A basis for the space of theories}\label{Sec:BasisSpaceOfTheories}

Note that as the conditions on $E$ and $\omega$ are linear, the space of partition functions given
two groups $\H \subset G$ carries a linear structure. Thus we can parametrise it by giving an
explicit basis for the space of distributions $E$ and $\omega$.

The distribution $\omega$ is a class function, thus the characters form a basis, and for compact groups it can be expanded as

\be\label{eq-faceDistrib}
\omega(g) = \sum_{\rho} \dim(\rho) \tilde{\omega}^\rho \tr_{\rho}\left(D_\rho(g)\right),
\ee
with $D_\rho(g)$ the representation matrix of $g$ in the unitary irrep $\rho$.

We will often set $\omega = \delta_G$, or $\tilde{\omega}^\rho = 1$ in which case the space of theories is simply parametrized by the functions $E$ satisfying the conditions of definition \ref{def-HolSFMNoBo}.

$E$ encodes the analogue of the simplicity constraints for the spin foam model at hand. The presence of the delta function on the face ensures that the product of group elements around the face, $g_f$, is flat. This is however not the usual holonomy around the face $\tilde{g}_f$, but $\tilde{g}_f$ interwoven with the $g_{ef}$. If we force $g_{ef} = 1$ we have $g_f = \tilde{g}_f$, and obtain a theory of flat connections. The presence of $g_{ef}$ and functions $E$ that allow them to differ from the identity thus relaxes the constraints on flatness.

Thus they exactly play the role of the simplicity constraints in ordinary spin foam models. We will
see the precise relationship between the simplicity constraints in the usual operator spin foam
models and the functions $E$ in the next section. We call this space of simplicity functions
$\EE(G,\H)$.

We can given an explicit basis for this space by expanding the functions in terms of the
irreducible unitary representations of $G$ and $\H$, which we denote $\rho$ and $k$ respectively.
We have chosen for every irreducible representation some specific
realization.
In the space $Hom_{\H}(\rho, k)$ we can introduce scalar product by
\begin{equation}\label{eq-GenInNormalisation}
 Hom_{\H}(\rho, k)\times Hom_{\H}(\rho, k)\ni(I,I')\rightarrow
\langle I,I'\rangle_{Hom_{\H}(\rho, k)} 1_k=I^\dagger I'\ .
\end{equation}
For every $\rho$ we have an antiunitary, group covariant map
\begin{equation}
 J_\rho\colon \rho\rightarrow\bar{\rho}
\end{equation}
If $\rho$ and $\bar{\rho}$ are distinct we assume that $J_{\bar{\rho}}=J_\rho^{-1}$.
If $\rho=\bar{\rho}$ then we may assume in addition that $J_\rho^2=s_\rho 1_\rho$ (where $s_\rho=\pm
1$).

Similarly, for every $k$ we have antiunitary map
\begin{equation}
 J_k\colon k\rightarrow\bar{k}
\end{equation}
If $k$ and $\bar{k}$ are distinct we assume that
$J_{\bar{k}}=J_k^{-1}$.
If $k=\bar{k}$ then we may assume in addition that $J_k^2=s_k 1_k$ (where $s_k=\pm 1$).

We can thus define the antiunitary (in the scalar product \eqref{eq-GenInNormalisation}) map
\begin{equation}
 M_{\rho,k}\colon Hom_{\H}(\rho, k),\rightarrow Hom_{\H}(\bar{\rho}, \bar{k}),\quad
M_{\rho,k}(I)=J_\rho I J_k^\dagger
\end{equation}

Here $J^\dagger$ is defined by

\begin{equation}
 \langle \cdot , J^\dagger \cdot \rangle = \overline{\langle J \cdot , \cdot \rangle}.
\end{equation}

In the case when $\rho=\bar{\rho}$ and $k=\bar{k}$
\begin{equation}
 M_{\rho,k}^2=\underbrace{s_ks_\rho}_{\pm 1} 1_{Hom_{\H}(\rho, k)}
\end{equation}

Let us define
\begin{equation}
 \tilde{D}(g)_{\rho,k}\colon Hom_{\H}(\rho, k)\rightarrow Hom_{\H}(\rho, k)
\end{equation}
by the matrix elements
\begin{equation}
 \langle I, \tilde{D}(g)_{\rho,k}I'\rangle=\tr I^\dagger D_\rho(g) I'\ .
\end{equation}
We can then expand $E$ as such:
\be
E(g) = \sum_{\rho, k} \dim(\rho) \tr_{Hom_{\H}(\rho, k)} e^\rho_k
\tilde{D}_{\rho,k}(g^{-1})
\ee
where $e^\rho_k\colon Hom_{\H}(\rho, k)\rightarrow Hom_{\H}(\rho, k)$.
$\EE(G,\H)$ can then be parametrized through $e^\rho_{k}$.

We will need the following set of useful relations satisfied by $\tilde{D}$. We see that
\begin{equation}
\begin{split}
\langle I, \tilde{D}_{\rho,k}(g^{-1})I'\rangle&=
\tr I^\dagger D_\rho^\dagger(g) I'=
\overline{\tr (I')^\dagger D_\rho(g) I}=\\
&=\overline{\langle I', \tilde{D}_{\rho,k}(g)I\rangle}
=\langle \tilde{D}_{\rho,k}(g)I,I'\rangle
\end{split}
\end{equation}
thus $\tilde{D}_{\rho,k}(g^{-1})=\tilde{D}_{\rho,k}(g)^\dagger$.

Similarly
\begin{equation}
\begin{split}
 &\langle I, M_{\rho,k}^\dagger\tilde{D}_{\bar{\rho},\bar{k}}(g)M_{\rho,k}(I')\rangle
 =\overline{\langle M_{\rho,k}(I), \tilde{D}_{\bar{\rho},\bar{k}}(g)M_{\rho,k}(I')\rangle}\\
&=\overline{\tr J_k\ I^\dagger\ J_\rho^\dagger J_\rho\
D_\rho(g)\ J_\rho^\dagger J_\rho\ I'\ J_k^\dagger}
=\langle I, \tilde{D}_{\rho,k} I'\rangle
\end{split}
\end{equation}
where we used identity valid for any antiunitary $J$ and linear $A$
\begin{equation}
 \tr J A J^\dagger=\overline{\tr A}
\end{equation}
Thus
$\tilde{D}_{\rho,k}(g)=M_{\rho,k}^\dagger\tilde{D}_{\bar{\rho},\bar{k}}(g)M_{\rho,k}$.

The $e^\rho_{k}$ are not completely free, but are restricted by the condition that $E(g) =
E(g^{-1})$. This implies that they have to satisfy a set of equations relating the coefficients for
complex conjugate representations. By definition we have that
\begin{equation}
\tr e^{\rho}_{k} \tilde{D}_{\rho,k}(g^{-1})=\overline{\tr
{e^{\rho}_{k}}^\dagger \tilde{D}_{\rho,k}(g)}
=\overline{\tr {e^{\rho}_{k}}^\dagger
M_{\rho,k}^\dagger\tilde{D}_{\bar{\rho},\bar{k}}(g)M_{\rho,k}}
=\tr M_{\rho,k}{e^{\rho}_{k}}^\dagger
M_{\rho,k}^\dagger\tilde{D}_{\bar{\rho},\bar{k}}(g)
\end{equation}
The condition for the E function reads\footnote{Matrix elements of $\tilde{D}_{\rho,k}$ satisfy
\begin{equation}
 d_\rho \int dg \langle I_1, \tilde{D}_{\rho,k}(g^{-1})I_2\rangle \langle I_3,
\tilde{D}_{\rho,k}(g)I_4\rangle =\delta_{\rho,\rho'}\delta_{k,k'} d_k
\langle I_1, I_4\rangle \langle I_3, I_2\rangle
\end{equation}
thus the basis form a set of independent functions of $g$.}
\begin{equation}
 e^{\bar{\rho}}_{\bar{k}}=M_{\rho,k}{e^{\rho}_k}^\dagger M_{\rho,k}^\dagger\ .
\end{equation}
Let us notice that if
$\rho\not=\bar{\rho}$ or $k\not=\bar{k}$ we can choose bases in such a way that $M_{\rho,k}$ acts by
complex conjugation, then
\begin{equation}\label{eq-simpleEcondition}
 e^{\bar{\rho}}_{\bar{k}}={e^{\rho}_k}^T
\end{equation}
The same is possible if $s_ks_\rho=1$ but not in the case when $s_ks_\rho=-1$\footnote{If the
degeneracy is $1$ then always $s_ks_\rho=1$ since the restriction of $J_\rho$ to $k$ is equal to
$J_k$}. From now on we will assume that $s_\rho s_k=1$ whenever $\rho=\bar{\rho}$ and $k=\bar{k}$.
In many models we will consider, $\rho$ and $k$ are indeed isomorphic to $\overline{\rho}$ and
$\overline{k}$, in which case the condition simply says that $e^\rho_{k}$ has to be symmetric.

Furthermore as
\begin{equation}
 \tr e^\rho_k\tilde{D}_{\rho,k}(g^{-1})=\tr e^\rho_k\tilde{D}_{\rho,k}(g)^\dagger
=\overline{\tr {e^\rho_k}^\dagger\tilde{D}_{\rho,k}(g)}
\end{equation}
we have a real holonomy spin foam model if
\be
{e^\rho_{k}} =  {e^\rho_{k}}^\dagger.
\ee
In the cases where we have \eqref{eq-simpleEcondition} this reads
\begin{equation}
 \overline{e^\rho_{k}} =  {e^{\bar{\rho}}_{\bar{k}}}.
\end{equation}
We can write these conditions explicitly by choosing an orthonormal basis in $Hom_\H(\rho,k)$
\begin{equation}
 I(\rho,k)_d
\end{equation}
satisfying conditions from above. In particular this means

\begin{equation}
I(\rho,k)_{d'}^\dagger I(\rho,k)_d = \delta_{dd'} \id_k
\end{equation}

Using formula $E(g)=\sum_\rho \dim(\rho)\ \tr e^{\bar{\rho}}_{\bar{k}}
D_{\rho,k}(g)$ and \eqref{eq-simpleEcondition}
the E function can be written as
\be
E(g) = \sum_{\rho, k} \dim(\rho) e^{\rho}_{k,dd'} \tr_\rho\left(D_\rho(g)\, I(\rho, k)_d
{I}^\dagger(\rho, k)_{d'}\right),
\ee

$\EE(G,\H)$ can then be parametrized through $e^\rho_{k,dd'}$. The index $d$ can be seen as a
degeneracy index for the case where the $\H$ reducible representation $\rho$ contains more than one
copy of the irreducible $k$.

\section{Boundaries and Hilbert spaces}\label{sec-BoundAndHilb}

We can now introduce a notion of boundary, which will lead us to a new, and more general notion of boundary Hilbert space.

\subsection{Boundaries}\label{sec-Boundaries}

We can now introduce and study boundaries into the formalism. To do so we chose an arbitrary graph $\Gamma$ in $\CC$, with edges $\Gamma_e \subset \CC_e$ and vertices $\Gamma_v \subset \CC_v$ subsets of the edges and vertices of $\CC$ respectively, as the boundary graph of $\CC$. We then write $\Gamma_{ev}$ for the set of pairs $v \in e$ in $\Gamma_v \times \Gamma_e$.

The partition function is then an element of the space $L^2(G^{|\Gamma_{ev}|})$ by dropping the integration over the group elements associated to these pairs:

\be\label{eq-PartFunc}
\ZZ^\Gamma (\CC)[g_{ev}] = \int \left(\prod_{e \subset f}  \dd g_{ef}\right) \left( \prod_{\substack{v \subset e \\ev \notin \Gamma_{ev}}} \dd g_{ev} \right) \left(\prod_{e \subset f} E(g_{ef})\right) \left(\prod_f \omega(g_f)\right).
\ee

This definition has the advantage that the inner product of the partition functions corresponds to the gluing along the graph. That is, for two complexes $\CC$, $\CC'$ with isomorphic boundary graphs we have:

\be\label{eq-gluing}
\left\la \overline{\ZZ^\Gamma (\CC)}, \ZZ^\Gamma (\CC')\right\ra^{\Gamma} = \ZZ (\CC \cup_{\Gamma} \CC')
\ee
where $\CC \cup_{\Gamma} \CC'$ indicates the 2-complex with $\Gamma$ in both complexes identified and now internal. This follows immediately from definitions. For a real spin foam model we of course can drop the complex conjugation. Note that the edges and vertices of $\Gamma$ become edges and vertices of $\CC \cup_{\Gamma} \CC'$. In particular these edges can be bivalent.

The integrand in \eqref{eq-PartFuncNoBo}, given by

\be\label{eq-Integrand}
\left(\prod_{e \subset f} E(g_{ef})\right) \left(\prod_f \omega(g_f)\right),
\ee
has the following symmetries:

\bea\label{eq-symmetries}
g_{ev} &\rightarrow& {h_e}^{-1}g_{ev}g_v\nn\\
g_{ef} &\rightarrow& {h_e}^{-1}g_{ef}h_e
\eea
for $h_e \in \H \subset G$ and $g_v \in G$.

Due to the symmetries of the integrand the partition function actually can be considered to live in
a smaller subspace of $L^2(G^{|\Gamma_{ev}|})$, that is, \be\ZZ \in \HH^\Gamma =
L^2\left(G^{|\Gamma_{ev}|}{\Bigl /}{\left(G^{|\Gamma_v|} \times {\H}^{|\Gamma_e|}\right)}\right),\ee
with the action of $(g_v, h_e) \in \left(G^{|\Gamma_v|} \times {\H}^{|\Gamma_e|}\right)$ on $g_{ve}$
by left and right multiplication: $(g_v, h_e) \triangleright g_{ve} = g_v g_{ve} h_e$.

We call this space the universal boundary space for the class of models $\EE(G,\H)$ and write
$\HH^\Gamma_{UBS}$.

This should be contrasted with the usual projected spin networks space \be\ZZ \in \HH^\Gamma_{PSN} =
L^2\left(G^{|\Gamma_{e}|}{\Bigl /}{\left(\H^{|\Gamma_v|}\right)}\right).\ee

\subsection{The spin network basis}\label{sec-SpinNetBasis}

We will now briefly give the spin network basis for the space $\HH_{UBS}^\Gamma$.

It is convenient to start with the basis for the larger space
$L^2(G^{|\Gamma_{ev}|})$. By the Peter-Weyl theorem this is given by the matrix
elements of representations, that is

\be|\rho,i,j\rangle =\sqrt{\dim(\rho)} D_{\rho}(g)_{ij} \q,\ee

with the dimension factor providing the correct normalisation,

\bea\langle\rho',i',j' |\rho,i,j\rangle &=& \sqrt{\dim(\rho)\dim(\rho')} \int_G \dd g \overline{D_{\rho'}(g)}_{i'j'} D_{\rho}(g)_{ij} \nn\\&=& \sqrt{\dim(\rho)\dim(\rho')} \int_G \dd g {D_{\rho'}(g^{-1})}_{j'i'} D_{\rho}(g)_{ij} \nn\\&=& \delta_{ii'} \delta_{jj'} \delta_{\rho\rho'}\q.\eea

A basis is thus simply given by the tensor product of basis elements

\be \bigotimes_{ev} |\rho_{ev},i_{ev},j_{ev}\rangle = \prod_{ev} \sqrt{\dim(\rho_{ev})}D_{\rho_{ev}}(g_{ev})_{i_{ev}j_{ev}} \in L^2(G^{|\Gamma_{ev}|})\q. \ee

$\HH_{UBS}^\Gamma$ is the subspace of states in $L^2(G^{|\Gamma_{ev}|})$ that are invariant under the action of the symmetries. In order to give a basis of this subspace it will actually be more convenient to use the orientation on the edges to introduce the oriented basis. For this we choose an arbitrary orientation for each edge. We will encode this by writing $(v,e,v') \in \Gamma$ for the oriented edge $e$ running from vertex $v'$ to $v$.

\be \bigotimes_{\substack{e\\(v',e,v) \in \Gamma}} \langle\rho_{ev'},i_{ev'},j_{ev'}| \tensor |\rho_{ev},i_{ev},j_{ev}\rangle \q.\ee

We now want to go to the $G$ invariant subspace of the $\rho_{ev}$ at the vertices, and the $\H$
invariant subspace at the edges. We begin by implementing the invariance under $\H$. To do so we
contract the inner indices at each oriented edge with an $\H$ covariant operator. These are
parametrized similarly to the $E$ functions, by the matrices $\Xi_{k,dd'}$ which can be contracted
with a basis of $\H$ invariant maps between $\rho$ and $\rho'$ giving

$$\Xi_{ii'} = \sum_{k,d,d',m}\Xi_{k,dd'} I(\rho,k)_{d,im}I(\rho',k)^\dagger_{d',mi'}.$$

Contracting these on the indices $i_{ev}$ in the middle of the edge gives

\bea|\rho_{ev},j_{ev}, \Xi_e\rangle &=& \prod_{\substack{e\\{(v',e,v) \subset \Gamma}}} \sqrt{\dim(\rho_{ev'})\dim(\rho_{ev})} \times\nn\\&&\times\; D_{\rho_{ev'}}(g_{ev'}^{-1})_{j_{ev'}i_{ev'}} \Xi_{e,i_{ev'},i_{ev}} D_{\rho_{ev}}(g_{ev})_{i_{ev}j_{ev}}\q.\eea

This is normalized as
\be\langle\rho'_{ev},j'_{ev}, \Xi'_e |\rho_{ev},j_{ev}, \Xi_e\rangle = \prod_{ev} \delta_{\rho_{ev}\rho'_{ev}} \delta_{j_{ev} j'_{ev}} \prod_e \tr {\Xi'_e}^{\dagger} \Xi_e \q . \ee
Note that $j_{ev}$ is in the dual to $\rho_{ev}$ if $(e,v) \subset \Gamma$, and in $\rho_{ev}$ directly if $(v,e) \subset \Gamma$.

For the case where there are no degeneracies $d,d'$ the coefficients $\Xi_{k,dd'}$ simplify to
$\Xi_{k}$. Thus we can directly work with the basis of $\H$ invariant operators $$\Xi'_{k,ii'} =
\frac1{\sqrt{\dim(k)}} I(\rho',k)_{im}I(\rho,k)^\dagger_{mi'}$$ labeled by $k$ with the property

\be \tr {\Xi'_{k}}^\dagger \Xi'_{k'} = \delta_{kk'}\q.\ee


We then obtain the states

\bea\label{snbasis1}
|\rho_{ev},j_{ev}, k_e\rangle &=& \prod_{\substack{e\\(v',e,v) \subset \Gamma}} \frac{\sqrt{\dim(\rho_{ev'})\dim(\rho_{ev})}}{\sqrt{\dim(k_{e})}} \times\\\nn&&\times D_{\rho_{ev'}}(g_{ev'}^{-1})_{j_{ev'}i_{ev'}} I(\rho_{ev'},k_{e})_{i_{ev'}m_{e}} I(\rho_{ev},k_{e})^\dagger_{m_{e}i_{ev}}D_{\rho_{ev}}(g_{ev})_{i_{ev}j_{ev}}\;,\eea

Which are normalized as

\be \langle\rho'_{ev},j'_{ev}, k'_e|\rho_{ev},j_{ev}, k_e\rangle = \prod_{ev}\delta_{\rho_{ev} \rho'_{ev}} \delta_{j_{ev} j'_{ev}} \prod_{e} \delta_{k_e k'_{e}} .\ee

To implement the $G$ invariance at the vertices we can simply contract with intertwiners $\eta_v \in \Inv(\otimes \rho^\star_{ev})$ where $\rho_{ev}^\star$ is the dual representation if $(v,e) \subset \Gamma$ and the usual representation if $(e,v) \subset \Gamma$. Contracting all these we obtain the state

\bea|\rho_{ev},\eta_v, \Xi_e\rangle &=& \prod_{v} \eta_{v,j_{ev},\dots} \prod_{\substack{e\\{(v',e,v) \subset \Gamma}}} \sqrt{\dim(\rho_{ev'})\dim(\rho_{ev})} \times\nn\\&&\times\; D_{\rho_{ev'}}(g_{ev'}^{-1})_{j_{ev'}i_{ev'}} \Xi_{e,i_{ev'},i_{ev}} D_{\rho_{ev}}(g_{ev})_{i_{ev}j_{ev}}\q.\eea

This is normalized as

\be\langle\rho'_{ev},\eta'_v, \Xi'_e |\rho_{ev},\eta_v, \Xi_e\rangle = \prod_{ev} \delta_{\rho_{ev}\rho'_{ev}} \prod_v \langle\eta'_v|\eta_v\rangle \prod_e \tr {\Xi'_e}^{\dagger} \Xi_e \q . \ee

For the case without degeneracies this again simplifies to the states

\bea\label{snbasis3}
|\rho_{ev},\eta_v, k_e\rangle &=& \prod_{v} \eta_{v,j_{ev},\dots} \prod_{\substack{e\\(v',e,v) \subset \Gamma}} \frac{\sqrt{\dim(\rho_{ev'})\dim(\rho_{ev})}}{\sqrt{\dim(k_{e})}} \times\\\nn&&\times D_{\rho_{ev'}}(g_{ev'}^{-1})_{j_{ev'}i_{ev'}} I(\rho_{ev'},k_{e})_{i_{ev'}m_{e}} I(\rho_{ev},k_{e})^\dagger_{m_{e}i_{ev}}D_{\rho_{ev}}(g_{ev})_{i_{ev}j_{ev}}\;,\eea

Which are normalized as

\be \langle\rho'_{ev},\eta'_v, k'_e|\rho_{ev},\eta_v, k_e\rangle = \prod_{ev}\delta_{\rho_{ev} \rho'_{ev}} \prod_v \langle\eta'_v|\eta_v\rangle \prod_{e} \delta_{k_e k'_{e}} .\ee

This should be contrasted to the basis of projected spin networks which is given in terms of $\H$
intertwiners $\iota$ as

\bea |k_{ev},\iota_v, \rho_e\rangle &=& \prod_v \iota_{v,m_{ev},\dots} \prod_e \sqrt{\dim(\rho_e)}\times\nn\\&&\times\; I(\rho_e,k_{ev})^\dagger_{m_{ev},j_{ev}} D_{\rho_e}(g_{vev'})_{j_{ev},j_{ev'}} I(\rho_e,k_{ev'})_{j_{ev'},m_{ev'}}\q,\eea

and is normalized as \be \langle k'_{ev},\iota'_v, \rho'_e|k_{ev},\iota_v, \rho_e\rangle = \prod_{ev}\delta_{k_{ev} k'_{ev}} \prod_v \langle\iota'_v|\iota_v\rangle \prod_{e} \delta_{\rho_e \rho'_{e}}\q.\ee

%
%
%
%
%
%

\subsection{Trimmed Complexes and Projected Spin Networks}\label{sec-TrimmedComplexes}

If the neighbourhood of the boundary of $\CC$ is of the form $\Gamma \times [0,1]$ we can make
contact to the projected spin network space $L^2 ({G}^{|\Gamma_e|}/{\H}^{|\Gamma_v|})$. In projected
spin networks the subgroup invariance is on the vertices of the boundary graph, rather than on the
edges. If the neighbourhood of $\Gamma$ is $\Gamma \times [0,1]$, every boundary vertex has an
associated internal edge $v \times [0,1]$, by ``splitting'' this associated internal edge we can
move the subgroup invariance to the boundary vertices. To do so we need a square root of the $E$
function, with the same subgroup covariance

\be\label{eq:F-sqrt-of-E}
E(g) = \int \dd g' F(g')F({g'}^{-1}g).
\ee

In terms of the basis coefficients this gives $f^\rho_{k} f^\rho_{k} =
e^\rho_{k}$. This means that whenever we have a term of the form $$\int \dd g
\dd g' f(gg')F(g)F(g')$$ with $g, g' \in G$ we can reparametrize with $\tilde{g}
= gg'$ and obtain 
$$\int \dd g \dd \tilde{g} f(\tilde{g})F(g)F(g^{-1}\tilde{g})
= \int \dd g f(g) E(g),$$
thus if the $E$ function defines a projector we obtain $E = F$.

While $F$ inherits the symmetries of $E$, that is, $F(hgh^{-1}) = F(g)$, for all
$h \in \H$, we
generally have that $F(g^{-1}) \neq F(g)$. Therefore the amplitudes constructed from $F$ will depend
on the orientations of the faces, and we will need to keep explicit track of the orientation of the
group elements associated to the edges in question. We do this by writing $g_{vev'} = g_{v'ev}^{-1}$
for $(vev') \subset f$ instead of $g_{ef}$, and we will split these as $g^f_{ve}g^f_{ev'}$.

Consider now the partition function associated to a ``cylindrical'' 2-complex $\CC = \Gamma \times [0,1]$ with boundary equal to two copies of $\Gamma$, $\Gamma^1$ and $\Gamma^2$. For simplicity for this section we will specialise to the case $\omega = \delta$. It is straightforward but notationally cumbersome to extend the discussion to the general case by introducing a square root of the face weight.
%
%

This partition function can then be factorized into the operators defined by

\bea
\mu^{\Gamma} [g_{ve}, \tilde{g}_{v'ev}] &=& \int \prod_{\substack{v \in \Gamma_v, e' \notin \Gamma_e \\ v \in e'}} \dd g_{ve'} \prod_{\substack{v \in \Gamma_v, e' \notin \Gamma_e \\ v \in e', e' \in f}} \dd g^f_{ve'} \prod_{\substack{e \in \Gamma_e\\ (vev') \subset f}} \dd g_{vev'} \times \nn\\&&\times \prod_{e \in \Gamma_e} \delta(g_{ve} g_{vev'} g_{ev'} g_{v'e'} g^f_{v'e'} \tilde{g}_{v'ev} g^f_{e''v} g_{e''v}) E(g_{vev'}) F(g^f_{v'e'}) F(g^f_{e''v})
\eea
where $(e''v)$ $(ve)$, $(ev')$, $(v'e')$ and $(vev') \subset f$, and $e \in \Gamma_e$. $e''$ and $e'$ are the edges $v \times [0,1]$, and $v' \times [0,1]$. The group elements in the delta function are those corresponding to the half of the face $f$ near the boundary edge $e$. Note that the group elements $\tilde{g}_{v'ev}$ are reversely oriented with respect to the boundary, they are on the ``opposite side'' of the half face in the delta.

$\mu$ defines a map from the projected spin network space $$\HH^\Gamma_{PSN} = L^2
({G}^{|\Gamma_e|}/{\H}^{|\Gamma_v|})$$ to the universal boundary space $$\HH^\Gamma_{UBS} =
L^2\left(G^{|\Gamma_{ev}|}{\Bigl /}{\left(G^{|\Gamma_v|} \times {\H}^{|\Gamma_e|}\right)}\right)$$
via
\bea
\psi_{UBS}(\{g_{ve}\})=\int \prod_{e\in \Gamma}  d\tilde g_{v'ev} \, \mu^{\Gamma} [g_{ve}, \tilde{g}_{v'ev}]  \psi_{PSN}( \{\tilde g_{v'ev}\}) \q .
\eea

This is constructed such that we have

\be\label{eq-mu-temporal}
\ZZ^{\Gamma^1 + \Gamma^2}(\Gamma \times [0,1])[g^1_{ve},g^2_{ve}] = \int \dd g^1_{vev'}\dd g^2_{vev'} \mu^{\Gamma^1}[g^1_{ve},g^1_{vev'}] {\mu^{\Gamma^2}}[g^2_{ve},g^2_{vev'}] \prod_{e\in \Gamma} \delta(g^1_{vev'} g^2_{v'ev}).
\ee

This can be seen by explicit calculation, however, these calculations are
greatly faciliated by the graphical notation we will introduce in the next
section, and we will illustrate them using examples there.

In the spin network basis for the case without degeneracies the $\mu$ map can be expressed as

\begin{equation}
\langle\rho_{ev},\eta_v,k_e|\mu|k_{ev},\iota_v, \rho_e\rangle = \prod_{ev} \delta_{\rho_{ev}\rho_e} \prod_v \big\langle\eta_v\big| \bigotimes_{e \ni v} I(\rho_{e}, k_{ev}) f^{\rho_{e}}_{k_{ev}}\big|\iota_v\big\rangle \;  \prod_e \sqrt{\frac{\dim(k_e)}{\dim(\rho_e)}}e^{\rho_e}_{k_e}.
\end{equation}


In general, if the boundary of the 2-complex is of the form $\Gamma \times [0,1]$ we can factorize its spin foam amplitude into the amplitude on the ``trimmed complex'' $\CC_t$, and $\mu$ for the boundary graph. The trimmed complex is the complex with ``half of the boundary faces taken off'', or, more technically, where we consider the boundary edges and vertices not to be part of the edge set and vertex set of the 2-complex but to be in a seperate set of boundary edges. We thus have $\CC_v$, $\CC_e$, $\CC_f$, $\Gamma_v$, $\Gamma_e$ as separate spaces, however, still with adjacency relations and orientations amongst each other as before.

\be
\ZZ^\Gamma(\CC) = \mu^\Gamma \tilde{\ZZ}^\Gamma(\CC_t)
\ee

with

\bea\label{eq-TrimmedPartFunc}
\tilde{\ZZ}^\Gamma (\CC_t)[g_{ef}] &=& \int \left(\prod_{e \subset f}  \dd g_{ef}\right) \left( \prod_{\substack{v \subset e \\e \in \CC_e}} \dd g_{ev} \right) \left(\prod_f \delta(g_f)\right)\times\nn\\&&\times \left(\prod_{\substack{e \in \CC_e, \subset f\\ \nexists v \in e, \in \Gamma_v}} E(g_{ef})\right)  \left(\prod_{\substack{e \in \CC_e, \subset f\\ \exists v \in e, \in \Gamma_v }} F(g_{ef})\right)
\eea

Note that this crucially depends on the orientations of the faces touching the boundary, and the composition of amplitudes only has the natural interpretation in terms of combining complexes if the orientations match up.

We see that $\mu$ embedds the projected spin network state space into the, in some sense larger, universal boundary space defined above, and clearly $\ZZ^\Gamma(\CC)$ lives in the image of $\mu^\Gamma$, thus we can equally well see the partition function as an element of the projected spin network space.

\subsubsection{Subgroup spin networks}

$\mu$ will generically have a non-zero kernel depending on $E$ or $F$ respectively. Thus we can
actually see $\tilde{\ZZ}$ as living in the coimage of $\mu$, which will allow us to go to subspaces
of projected spin networks, for example $\H$ spin networks, as was discussed in
\cite{Dupuis:2010jn}. This the case for example in the EPRL spin foam model.

We can realize this restriction to subgroup spin networks explicitly if there is an $\omega'$ with the properties of $\omega$ such that the $E$ function satisfies

\be
E(g)\omega(gg'g''g''')E(g'') = \int_\H dh \omega'(g'gh)E(g)\omega'(g''g'''h^{-1})E(g'').
\ee

We can then glue via subgroup integrations on the boundary edges. Again, it is easy to see that this leads to the correct gluing using the graphical notation in the next section. We assume that every face has at most one boundary edge and replace $\omega$ on those faces with $\omega'$, as well as reducing the boundary group element to live in the subgroup. The partition function then becomes

\bea\label{eq-SubsetPartFunc}
\tilde{\ZZ}^\Gamma (\CC_t)[h_{e}] &=& \int \left(\prod_{e \subset f}  \dd g_{ef}\right) \left( \prod_{\substack{v \subset e \\e \in \CC_e}} \dd g_{ev} \right) \left(\prod_{f} \omega^\star(g_f)\right)\times\nn\\&&\times \left(\prod_{\substack{e \subset f\\ e \notin \Gamma_{e}}} E(g_{ef})\right) \left(\prod_{\substack{e \subset f\\ e \in \Gamma_e}} E(h_e g_{ef})\right),
\eea

where $\omega^\star$ is $\omega$ if the face does not contain a boundary edge, and $\omega'$ if it does.

\subsection{A graphical notation}

It is illuminating to illustrate the structure of the convolutions in the partition function using a graphical notation. This will allow us to explicitly keep track of the way group elements in different faces are identified. The graphical notation will have three ingredients, corresponding to the face amplitude $\omega$, the insertion of $E$ around a face, and the insertion and integration of the $g_{ev}$.

\begin{figure}[htbp]
 \centering
 \includegraphics[scale=.5]{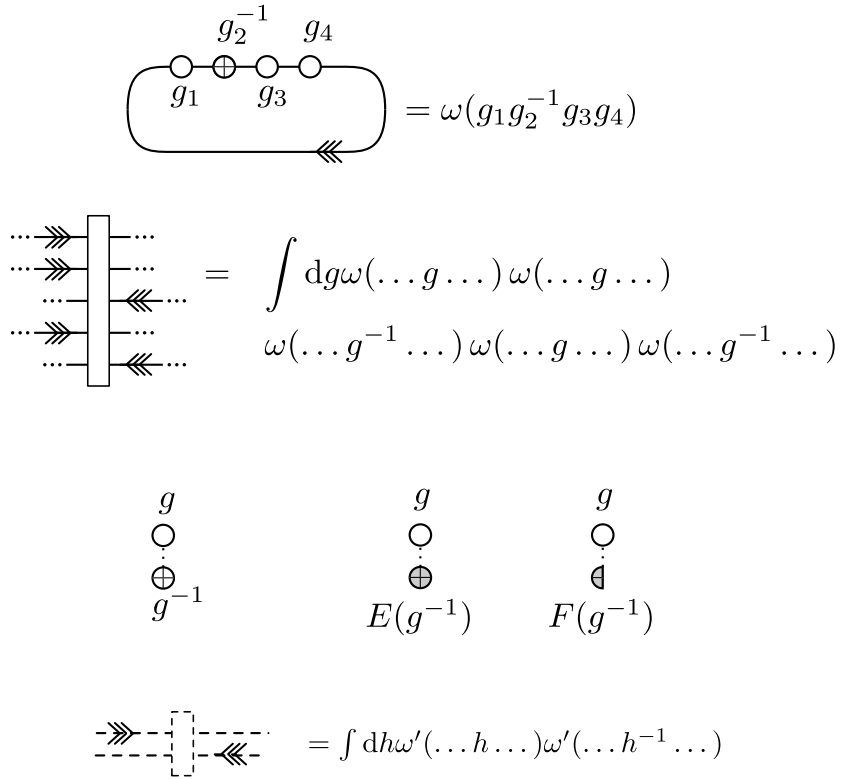}
 \caption{Ingredients of the graphical notation.}
 \label{fig-Ingredients}
\end{figure}

We will indicate $\omega$ by a solid line. White circles crossed by the line indicate group elements that are multiplied together to form the argument of $\omega$. The fins of the line indicate the order in which the inserted group elements should be multiplied. Two circles joined by a dotted line indicates the same group element. If one of them is crossed the group elements should be inverse of each other. The $E$ function is indicated by a grey circle, the cross indicates that the argument in the $E$ function should be the inverse of that it is linked to. We will represent its convolution square root $F$ with a grey half circle. A white box indicates an insertion of oriented group elements into the lines that pass through it. This is given in Figure \ref{fig-Ingredients}.

\begin{figure}[htbp]
 \centering
 \includegraphics[scale=.5]{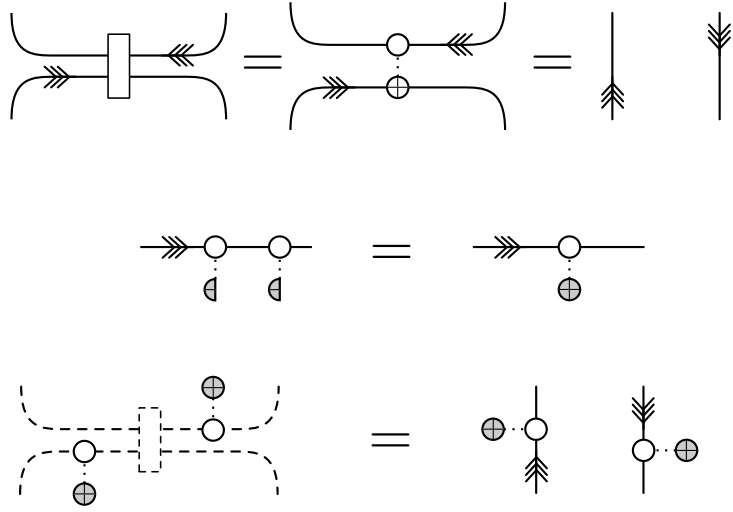}
 \caption{Some possible relations among the ingredients. The upper identity holds for $\omega = \delta$, the second is a consequence of the definition of $F$, the third is the subgroup property.}
 \label{fig-Relations}
\end{figure}

\begin{figure}[htbp]
 \centering
 \includegraphics[scale=.3]{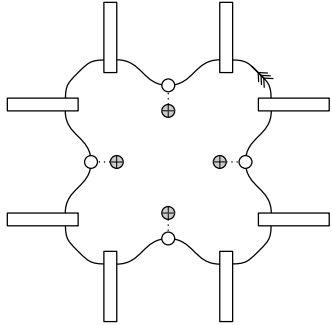}
 \caption{The structure of a face.}
 \label{fig-Face}
\end{figure}

\begin{figure}[htbp]
 \centering
 \includegraphics[scale=.3]{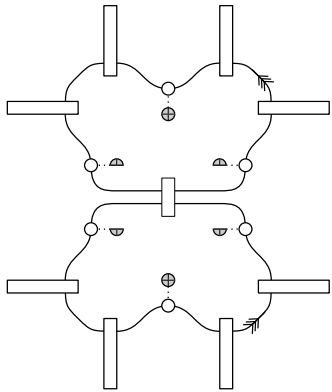}
 \caption{Two half faces convoluted.}
 \label{fig-HalfFace}
\end{figure}

Using these ingredients we can represent the structure of a face containing four edges as in Figure \ref{fig-Face}. The structure of two half faces, and their composition in the projected spin network space, is given in Figure \ref{fig-HalfFace}. It is now a straightforward application of the relations illustrated in Figure \ref{fig-Relations} to see that the two half faces do indeed compose to a full face.

We can write out the composition illustrated in \ref{fig-HalfFace}. Let that face be $f=(012345670)$, with $0,2,4,6 \in \CC_v$ and $1,3,5,7 \in \CC_e$. Taking care of the orientations in the delta function and the $F$, the integrand for a face on the right hand side reads

\begin{eqnarray}
&\delta(g_{01} g_{012} g_{12} g_{23} g^f_{23} \tilde{g}_{210} g^f_{70} g_{70}) \delta(g_{45} g_{456} g_{56} g_{67} g^f_{67} \tilde{g}_{654} g^f_{34} g_{34})\times&\nn\\&\times E(g_{012}) F(g^f_{23}) F(g^f_{70})E(g_{456}) F(g^f_{67}) F(g^f_{34})\delta(\tilde{g}_{210}\tilde{g}_{674}).&
\end{eqnarray}

After combining the delta functions we obtain:

\begin{eqnarray}
&\delta(g_{01} g_{012} g_{12} g_{23} g_{23}^f g_{34}^f g_{34} g_{45} g_{456} g_{56} g_{67} g^f_{67} g^f_{70} g_{70}) \times&\nn\\&\times E(g_{012}) F(g^f_{23}) F(g^f_{70}) E(g_{456}) F(g^f_{67}) F(g^f_{34}).&
\end{eqnarray}

Which, with equation \eqref{eq:F-sqrt-of-E}, gives the amplitude for a face in the normal partition function, pictorial represented in figure \ref{fig-Face},

\begin{eqnarray}
&\delta(g_{01} g_{012} g_{12} g_{23} g_{234} g_{34} g_{45} g_{456} g_{56} g_{67} g_{670} g_{70}) \times&\nn\\&\times E(g_{012}) E(g_{234}) E(g_{456}) E(g_{670}).&
\end{eqnarray}

\begin{figure}[htbp]
 \centering
 \includegraphics[scale=.3]{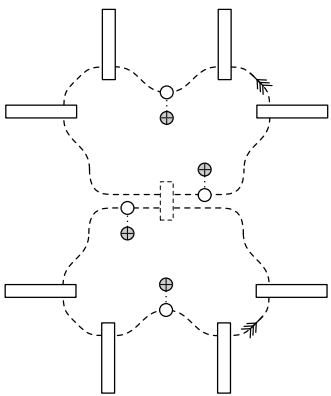}
 \caption{Two half faces convoluted using the subgroup property.}
 \label{fig-SubgroupFace}
\end{figure}

If we further have the subgroup property on $E$ we can replace the convolution in the $G$ with a
convolution in $\H$, thus reducing to the sub group spinnetworks. This is illustrated in Figure
\ref{fig-SubgroupFace}.

\section{Gluing of 2-complexes}\label{glue glue}

\begin{figure}[htbp]
 \centering
 \includegraphics[scale=.3]{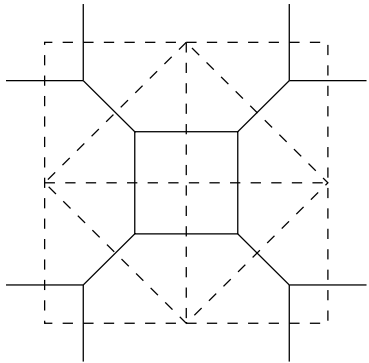}
 \caption{A part of a triangulation (dashed lines), and its dual (solid lines).}
 \label{fig-Triangulation}
\end{figure}

We will now show how the different Hilbertspaces discussed above give correspond to different gluing operations on the 2-complexes. To simplify illustrations we will focus on the case of the 2-complex dual to a triangulated surface, and various related 2-complexes. In that case the dual 2-complex has trivalent vertices and bivalent edges, see for example Figure \ref{fig-Triangulation}. The 2-complex contains a central face with four edges, which has the structure illustrated in Figure \ref{fig-Face}.

\begin{figure}[htbp]
 \centering
 \includegraphics[scale=.5]{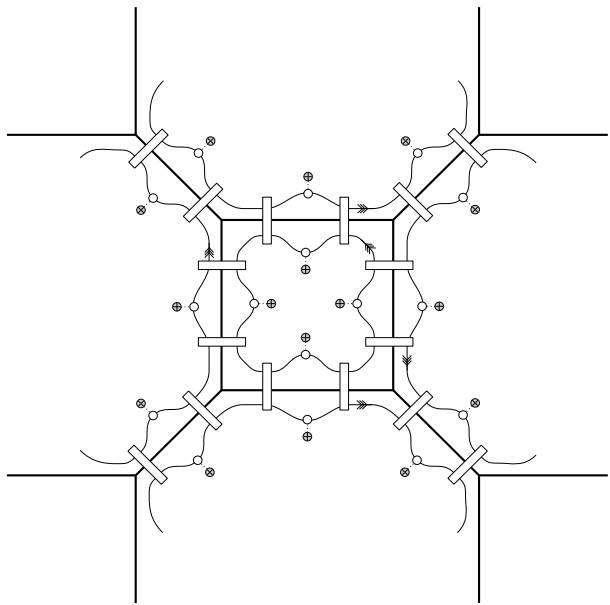}
 \caption{The part of the partition function corresponding to the part of the triangulation shown in Figure \ref{fig-Triangulation}.}
 \label{fig-Tri-Amplitude}
\end{figure}

The partition function for this part of the amplitude is given in \ref{fig-Tri-Amplitude}. This ampltiude can be obtained by various gluings from different building blocks, depending on how we take it apart.

\subsection{Face to face: the universal boundary space}\label{sec-FaceToFace}

The first gluing is simply that corresponding the universal boundary space. As noted above, arbitrary 2-complexes can be glued along arbitrary edges. Thus we can in particular simply take the faces of the 2-complex as individual partition functions. These partition functions for single faces were introduced in the companion paper as effective face weights $\omega_f$.

\be\label{eq-effectiveFaceWeights}
\omega_f=\ZZ(f)
\ee

\begin{figure}[htbp]
 \centering
 \includegraphics[scale=.3]{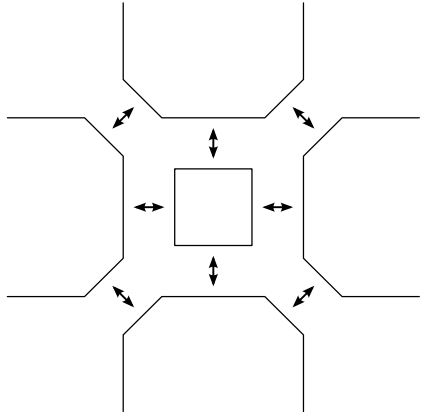}
 \caption{The composition of the 2-complex from faces. Each small double arrow indicates a composition in the universal boundary space associated to one edge.}
 \label{fig-Gluing-Omega}
\end{figure}

We obtain one such $\omega_f$ per type of face. In particular we have only one such effective face weight for a regular complex. These can then be composed simply by equation \eqref{eq-gluing} to yield arbitrary 2-complexes. This is sketched for the complex above in Figure \ref{fig-Gluing-Omega}.

Note that the composition in $\HH^\Gamma_{UBS}$ as given by \eqref{eq-gluing} can be localize to a subset of the boundary graph. That is, given $\Gamma \in \CC$ and $\Gamma' \in \CC'$ and a graph $\Gamma^b$ that is a subgraph of $\Gamma$ and $\Gamma'$, we can treat only the common sub graph as boundary, and glue amongst it, yielding,
\be\label{eq-partial-boundary-trace}
\tr^{\Gamma^b} \ZZ^\Gamma(\CC) \ZZ^{\Gamma'}(\CC') = \ZZ^{\Gamma\cup_{\Gamma^b}\Gamma'}(\CC\cup_{\Gamma_b}\CC'),
\ee
where $\Gamma\cup_{\Gamma^b}\Gamma'$ is the graph obtained by identifying $\Gamma^b \in \Gamma, \Gamma'$ and then deleting the identified graph, and $\tr^{\Gamma^b}$ indicates integrating the elements of $\HH_{UBS}^{\Gamma^b} \tensor {\HH'}_{UBS}^{\Gamma^b}$ against the element $\prod_{ev}\delta(g_{ev} {g'_{ev}}^{-1})$.

Note that due to gauge invariance at the two-valent vertices $\omega_f$ depends only on as many variables as $f$ has edges. It will be convenient to also introduce $\omega'_f$ such that $$\omega_f(g_{ve},g_{ev'},g_{v'e'},g_{e'v''},\dots) = \omega'_f(g_{ve}g_{ev'},g_{v'e'}g_{e'v''},\dots).$$

\subsubsection{A special case: Wedges to wedges}

\begin{figure}[htbp]
 \centering
 \includegraphics[scale=.3]{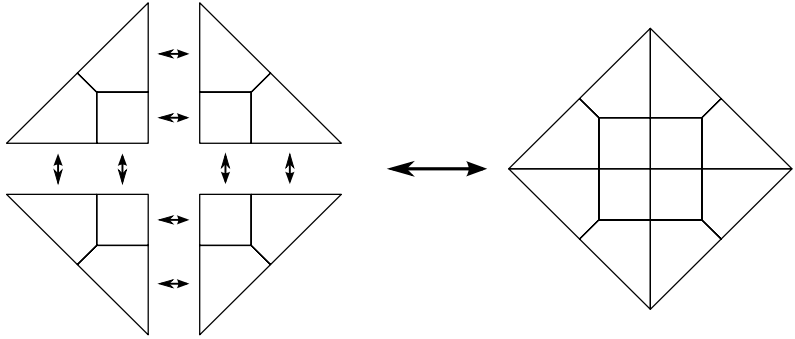}
 \caption{The composition of the 2-complex from the content of the simplex. Each small double arrow indicates a composition in the universal boundary space associated to one edge.}
 \label{fig-Gluing-F2F}
\end{figure}

In the case of the complex dual to a triangulation each type of face can ocurr, thus effective face weights are not a convenient choice for constructing theories. However, we can construct a second kind of dual 2-complex $\CC'$ constructed from so called wedges. That is, we take as faces the intersections of the faces of the dual 2-complex and the simplices. In our 2-dimensional example this means that each triangle now contains three such wedges. These can then be composed in the universal boundary space again. The advantage is that now we only need one type of amplitude that we are gluing, the content of a simplex. This is illustrated in Figure \ref{fig-Gluing-F2F}, the new complex $\CC'$ is on the right.

Calling the complex of wedges $\sigma^*$, and using the trace from equation \eqref{eq-partial-boundary-trace} we can thus write the entire partition function for a dual complex made from wedges,

\be
\ZZ(\CC')=\tr^\Gamma \bigotimes_v \ZZ(\sigma^*)
\ee

\subsection{Half face to half face: the spin network spaces}

As noted above, the gluing of two trimmed partition functions in $\HH^\Gamma_{PSN}$ also generates a natural composition,

\be\label{eq-gluingPSN}
\left\la \overline{\tilde{\ZZ}^\Gamma (\CC)}, \ZZ^\Gamma (\CC')\right\ra^{\Gamma}_{PSN} = \tilde{\ZZ} (\CC \tilde{\cup}_{\Gamma} \CC').
\ee

Due to the topological restrictions near the boundary required for trimming the partition function, all edges of $\Gamma$ in $\CC \cup_{\Gamma} \CC'$ are bivalent, and $\CC \tilde{\cup}_{\Gamma} \CC'$ is the 2-complex obtained by identifying the two copies of $\Gamma$ and then erasing the bivalent edges in $\Gamma_e$ and then the bivalent vertices in $\Gamma_v$. This is the composition usually done in spin foam models.

\begin{figure}[htbp]
 \centering
 \includegraphics[scale=.25]{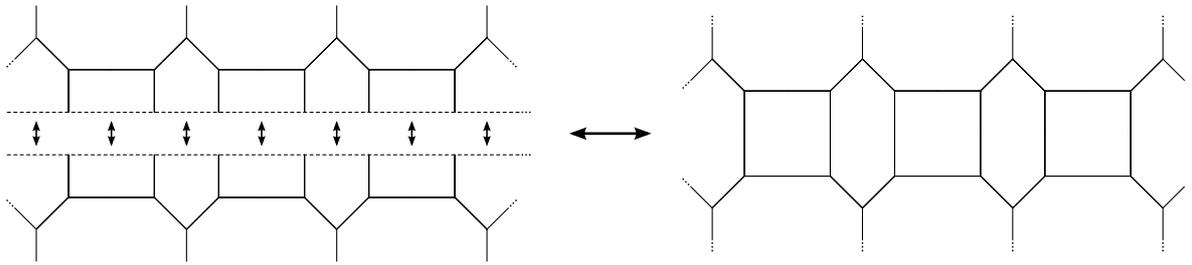}
 \caption{The composition of the 2-complex from half faces. Each small double arrow indicates a composition in the projected spin network space associated to one edge.}
 \label{fig-Gluing-HalfFace}
\end{figure}

For the case of a 2-complex dual to a triangulation this is a natural type of gluing, an example is illustrated in Figure \ref{fig-Gluing-HalfFace}. The half faces and composition on the left hand side are exactly those illustrated in Figure \ref{fig-HalfFace}.

\subsection{Half faces around a face}

We can generalize the above gluing by using the partition function with $\delta$ functions on the half faces, and gluing around a face amplitude $\omega$ in the following sense.

\begin{figure}[htbp]
 \centering
 \includegraphics[scale=.25]{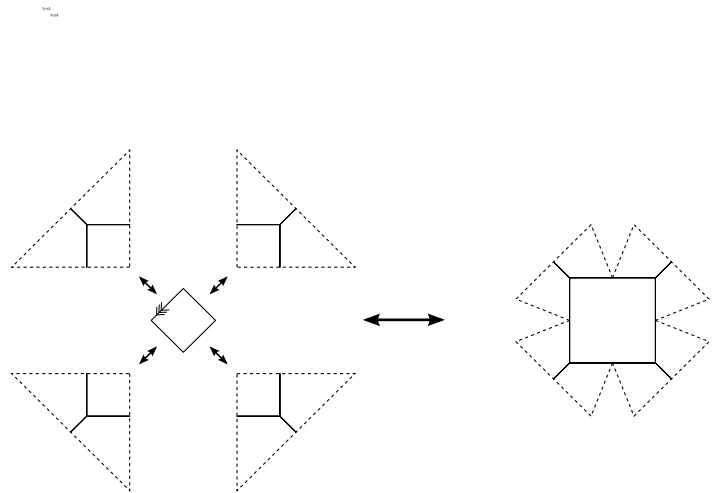}
 \caption{The composition of half wedges around a face.}
 \label{fig-Gluing-Wedges}
\end{figure}

For a set of boundary edges $e_a, a = 1 \dots n$ in a set of trimmed complexes $\tilde{\CC}_a$, we can form the new partition function on a trimmed complex with an interior face $f=(v,e_1,v', e_2, \dots, e_n,v)$

\begin{equation}\label{eq-PSN_face_gluing}
\tilde\ZZ\left(\bigcup_{f} \tilde{\CC}_a\right) = \int \dd g_{e_a} \omega\left(\prod_a g_{e_a}\right) \prod_a \tilde{\ZZ}_a(\tilde{\CC}_a).
\end{equation}

This is illustrated in Figure \ref{fig-Gluing-Wedges}.

In this way we can parametrize the partition function of the dual complex of a triangulation by the complex of trimmed wedges in a simplex and the face weight.

\section{Standard spin foam models in holonomy language.}\label{sec-StandardSFM}

In this section we show that the BC, EPRL and FK model can be expressed in the language above. To do so we will relate the Holonomy model defined above to the operator models of \cite{Bahr:2010bs}. We can then give natural generalisations of the BC, EPRL and FK model to arbitrary, and, in particular, finite groups.

\subsection{From operators to holonomies}

We can relate the OSFM to the ones by insertion of the group integrations $g_{ef}$ at each pair of edge and face by using the orthogonality of group elements:

\be\label{eq-orthogonality}
\dim(\rho) \int \dd g \,{{D_\rho}^b_a (g^{-1})} {D_{\rho'}}_{b'}^{a'} (g) = \delta_{a}^{a'}\delta_{b'}^{b}\delta_{\rho\rho'}\,.
\ee

\begin{figure}[htp]
 \centering
 \includegraphics[scale=.5]{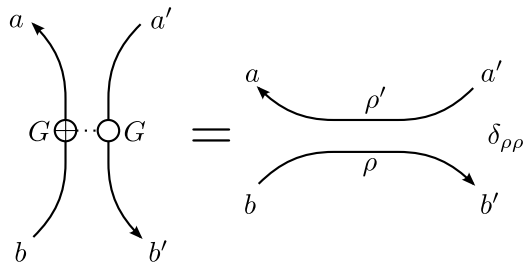}
 \caption{Orthogonality in graphical notation.}
 \label{fig-Ortho}
\end{figure}

The structure of the manipulation is easiest to see using an extension of the graphical notation. We will decorate the lines with irreducible representations. We can then break them into segments with indicies living in the representation with arrows indicating the ingoing and outgoing indices. In Figure \ref{fig-Ortho} we use this calculus to express equation \eqref{eq-orthogonality}. A line joining two objects indicates the composition of tensors. A line coming into a tensor indicates an index in a representation space (downstairs) and a line outgoing indicates an index in the dual space (upstairs).

We will start by deriving the holonomy formulation given in the previous
section, from the edge operator formalism of \cite{Bahr:2010bs}. There we have,
for a colouring of the faces by irreducible representations $\rho$, an operator
$P$ on each edge. These operators are then contracted according to the structure
of the 2-complex. For simplicity we will focus on the case where the edge
orientation and the face orientation agree. Then we have $P_e \in \Hom(\rho_f
\tens \rho_{f'} \dots \rho_{f''})$ where $f,f',f'' \ni e$. In all models studied
so far except the KKL model \cite{Kaminski:2009cc, Kaminski:2009fm} $P$ has an
additional factorisation property which implies that it can be expressed in the
following way:

\be
{P_e}^{(a)}_{(a')} = {P_{G}}^{(a)}_{(b)}\; \tilde{E}^{b_f}_{c_f} \tens \tilde{E}^{b_{f'}}_{c_{f'}} \tens \dots \tens \tilde{E}^{b_{f''}}_{c_{f''}}\; {P_{G}}^{(c)}_{(a')},
\ee

where $(a)$ is a multi index ranging over $a_f$, with $f \ni e$, and $P_{G}$ is the projector on the gauge invariant subspace. The $\tilde{E} \in L(\rho_f)$ are a set of one linear operator per representaton space which satisfy relations ensuring that the fiducial orientations do not enter in contracting the edge operators. These are not group covariant. We will usually supress dependence of $\tilde{E}$ on the representation space as it should be clear from context on which space it acts. If we write this projector explicitly as a gauge averaging we obtain the following:

\be
{P_e}^{(a)}_{(a')} = \int \dd g_{ve} \dd g_{ev}\; \bigotimes_{f\ni e} {D_{\rho_f}}_{b_f}^{a_f} (g_{ve})\; \tilde{E}^{b_f}_{c_f}\; {D_{\rho_f}}_{a'_f}^{c_f} (g_{ev}).
\ee

\begin{figure}[htp]
 \centering
 \includegraphics[scale=.5]{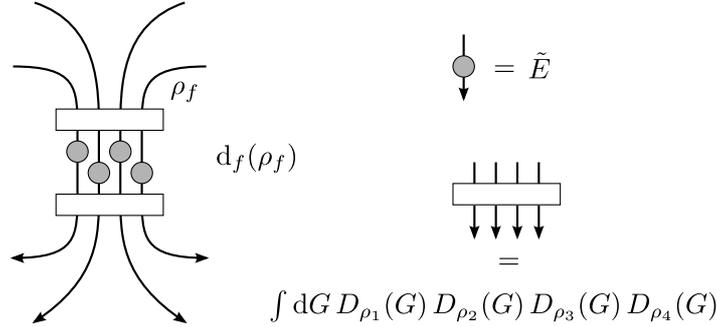}
 \caption{Factorisation of the edge operator in graphical notation.}
 \label{fig-FactorStruct}
\end{figure}

The graphical representation of the edge operator is given in Figure \ref{fig-FactorStruct}.

Inserting the resolution of the identity \eqref{eq-orthogonality}, and introducing a sum over representations, we obtain

\begin{eqnarray}
{P_e}^{(a)}_{(a')} = \int \dd g_{ve} \dd g_{ev'}& \displaystyle \prod_{f \ni e}
\dd g_{ef} &\bigotimes_{f \ni e} {D_{\rho_f}}_{b_f}^{a_f} (g_{ve})\;
{D_{\rho_f}}^{b_f}_{c_f} (g_{ef})\; {D_{\rho_f}}_{a_f}^{c_f} (g_{ev'})
\times\nn\\&&\times \sum_{\rho_{ef}} \dim(\rho_{ef})
{D_{\rho_{ef}}}^{c'_f}_{b'_f} (g^{-1}_{ef}) \; \tilde{E}^{b'_f}_{c'_f}.
\end{eqnarray}
with $\sum_{\rho_{ef}} \dim(\rho_{ef}){{D_{\rho_{ef}}}^{c'_f}_{b'_f}
(g_{ef}^{-1})} \; \tilde{E}^{b'_f}_{c'_f}$ being the function $E$. The
representation matrices around a face can be contracted to a character, and as
we decoupled the representation label on the operators $\tilde{E}$ the sum over
$\rho_f$ can be performed exactly to arrive at the distribution
\eqref{eq-faceDistrib} and we arrive at \eqref{eq-PartFuncNoBo}. Note that the
conditions on $\tilde{E}$ that imply independence of the fiducial orientations
now imply $E(g) = E(g^{-1})$ and the end result is indeed invariant under
reversing the orientations.

\begin{figure}[htp]
 \centering
 \includegraphics[scale=.5]{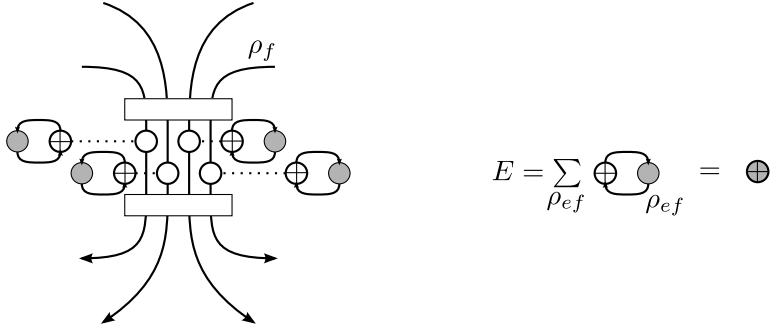}
 \caption{The E functions.}
 \label{fig-E-func}
\end{figure}

The graphical representation of this insertion of identities is given in Figure \ref{fig-E-func}.

\begin{figure}[htp]
 \centering
 \includegraphics[scale=.5]{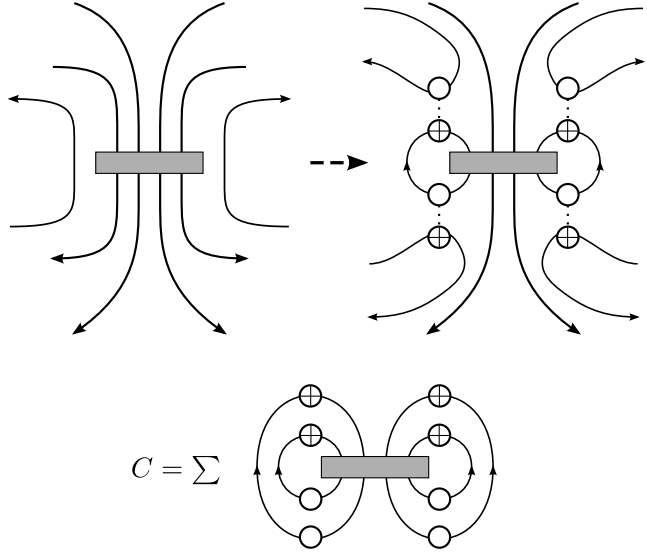}
 \caption{The C functions.}
 \label{fig-C-func}
\end{figure}

In the companion paper \cite{Comp} as well as in \cite{Bahr:2011aa} more general edge operators that do not have this factorisation property are also considered. Then there still is a holonomy formulation, given that the face amplitude is simply the dimension of $\rho_f$. We briefly recall this construction and illustrate it in the graphical calculus in Figure \ref{fig-C-func}.

The dimension as a face amplitude can be given by the trace of the identity operator, or a closed circle in graphical notation. The crucial ingredient is then a function on $n$ copies of the group, given simply by

\be\label{f1}
C(g_{ef}, \dots ,g_{ef'}) = \sum_{\substack{\rho_{ef}\\e\in f}} \left( \prod_{f\ni e} \dim(\rho_{ef}) {D_{\rho_{ef}}}_{b_f}^{a_f}(g_{ef})\right) P^{(b)}_{(a)}.
\ee

The reason for this formula is clear by Figure \ref{fig-C-func}. From this diagram we can also see that the $C$ functions are glued with delta functions. In particular the integrand is schematically given by \be \prod_e C(g^v_{ef}{g^{v'}_{ef}}^{-1}, \dots) \prod_v \prod_{(eve') \subset f} \delta(g^v_{ef}{g^v_{e'f}}^{-1}).\ee

In GFT language the $C$ are simply the propagators and $\prod_{(eve') \subset f} \delta(g^v_{ef}{g^v_{e'f}}^{-1})$ is the vertex function, or interaction, that glues them.

\subsection{The specific models for Spin(4)}

Having in hand the relationship to the operator formalism, we can now easily give the formulation of the various established spin foam models in the holonomy language. We begin with the BF, BC \cite{Barrett:1997gw} and EPRL model \cite{Engle:2007wy} which are quite straightforward, and then discuss the FK \cite{Freidel:2007py} and BO \cite{Baratin:2011hp,Baratin:2011tx} in the subsequent subsections.

\subsubsection{BF, Barrett-Crane and Engle-Pereira-Rovelli-Livine}

We specify now to $G = \Spin(4)$ and $\H = \SU(2) = \Spin(4)_{diag}$ the diagonal $\SU(2)$ subgroup,
with irreps labelled by $\rho$ and $k$ respectively. The operators $\tilde{E}$ in $L(\rho)$ for the
first set of models are given by the following:

\be
{\tilde{E}_{BF}} = \id,
\ee
for $BF$ theory,

\be
{\tilde{E}_{BC}} = \dd_e(\rho) \sum_{k} \delta(\rho,(k,k)) I(\rho,0) I(\rho,0)^\dagger,
\ee
for the Barrett Crane model, where $\dd_e(\rho)$ is an arbitrary edge meassure factor, and for the EPRL model \cite{Engle:2007uq,Engle:2007qf,Engle:2007wy} we have:

\be
{\tilde{E}_{EPRL}} = \dd_e(\rho) \sum_{k} \delta\left(\rho,\rho_\gamma(k)\right) I(\rho, k) {I(\rho,k)}^\dagger
\ee
where we write $\rho_\gamma(k) = \left(\frac{1+\gamma}{2} k, \frac{|1-\gamma|}{2}k\right)$.


We can then obtain the $E(g)$ by the formulas of the previous section. For the EPRL model the functions $E^\gamma_{EPRL}(g)$ is given by

\be
E^\gamma_{EPRL}(g) = \sum_{\rho, k} \dim(\rho) \dd_e(\rho) \delta\left(\rho,\rho_\gamma(k)\right) \tr_\rho\left(D_\rho(g)\, I(\rho,k) I(\rho,k)^\dagger\right),
\ee
and for Barrett-Crane we obtain

\be
E_{BC}(g) = \sum_{\rho, k, k'} \dim(\rho) \dd_e(\rho) \delta(k) \delta\left(\rho,(k',k')\right) \tr_\rho\left(D_\rho(g)\, I(\rho,k) I(\rho,k)^\dagger\right).
\ee

The $\SU(2)$ irrep $k'$ plays a very different role than in the EPRL model, namely it restricts the form of the irrep $\rho$ but does not appear in the injection maps $I(\rho,k)$. We can further simplify this by noting that

\bea
\tr_{(k,k)}\left(D_{(k,k)}(g)\, I({(k,k)},0) I({(k,k)},0)^\dagger\right) &=& I((k,k),0)^\dagger D_{(k,k)}(g) I((k,k),0)\nn\\ &=& \tr_k(g^+{g^-}^{-1}) \dim(k)^{-1},
\eea
with the $\Spin(4)$ element $g$ decomposing into the left and right $\SU(2)$ as $g = (g^+, g^-)$. Thus for choice $\dd_e(\rho) = 1$ we have simply

\be
E_{BC}(g) = \delta(g^+{g^-}^{-1}) = \delta_{\Spin(4)_{diag}}(g),
\ee
where, for a general subgroup $\H \subset G$ we write \be\delta_{\H}(g) = \int_{\H} \dd h
\delta(gh^{-1})\ee for the delta function that force a group element to lie in the subgroup.

Thus we arrive at a particularly simple form for the Barrett-Crane model as an integral over a product of $\SU(2)$ delta functions:

\be\label{eq-PartFuncBC-OnlyDeltas}
\ZZ_{BC}(\CC) = \int \left(\prod_{e \subset f}  \dd h^\pm_{ef}\right) \left( \prod_{v \subset e} \dd g^\pm_{ev} \right) \left(\prod_{e \subset f} \delta\left(g^+ \left(g^-\right)^{-1}\right)\right) \left(\prod_f \delta(g^+_f)\delta(g^-_f)\right).
\ee

For the FK model the $E$ function is most easily expressed in terms of coherent states, that is, the eigenstates of the Lie algebra generators. For an $\SU(2)$ representation labelled by the half integer  $k$ these are the states $\alpha(\nb)$ that satisfy $(\nb \cdot L) \alpha_k(\nb) = i k \alpha_k(\nb)$, where $\nb$ is a unit vector in $\R^3$. All $\alpha_k(\nb)$ for the same $\nb$ differ at most by a phase. The $E$ function for the FK model is given by

\be\label{eq-EfuncFK}
E_{FK}(g^+,g^-) = \sum_k \dim(k) \int \dd\nb \left( \alpha^\dagger_k(\nb)D_k(g^+)\alpha_k(\nb)\right)\overline{\left( \alpha^\dagger_k(\nb)D_k(g^-)\alpha_k(\nb)\right)},
\ee
which is well defined as the phases of $\alpha_k(\nb)$ and $\alpha^\dagger_k(\nb)$ cancel.

BF theory of course is simply given by setting \be E_{BF}(g) = \delta(g).\ee

The coefficients in the basis of section \ref{Sec:BasisSpaceOfTheories} for the various models then are:

\begin{itemize}
 \item BF: $e^{\rho}_k = 1$
 \item BC: $e^{\rho}_k = \dd_e(\rho) \delta_{k,0} \sum_{k'} \delta(\rho,(k',k'))$
 \item EPRL: $e^{\rho}_k = \dd_e(\rho) \delta\left(\left(\frac{1+\gamma}{2} k, \frac{|1-\gamma|}{2} k\right),\rho\right)$
\end{itemize}

We see from the conditions derived in section \ref{Sec:BasisSpaceOfTheories}
that for real $\dd_e(\rho)$ all these spin foam models are indeed real holonomy
spin foams.

\subsubsection{Freidel-Krasnov for $\gamma>1$}

The FK model is defined in terms of coherent states. For $\gamma < 1$ it is equivalent to the EPRL model. To obtain its basis coefficients for $\gamma > 1$ we need to work some more. In this section we use the shorthand $|m\rangle = \alpha(m)$.

The operator $\tilde{E}_\rho$ for the FK model is given by
\begin{equation}
 \tilde{E}_\rho=\dd_e(\rho)\int\frac{dm}{4\pi} |m\rangle^{2j^+} \langle
m|^{2j^+}\otimes
|\bar{m}\rangle^{2j^-}
\langle \bar{m}|^{2j^-}
\end{equation}
with the representations $\rho=(j^+,j^-)$ subject to constraints similar to those in the EPRL model.
Let
\begin{equation}
 R_\gamma=\left\{\left(\frac{\gamma+1}{2}
\tilde{k},\frac{\gamma-1}{2} {\tilde{k}}\right)\colon
\tilde{k}=0,\frac{1}{2},\ldots\right\}.
\end{equation}
We can write
\begin{equation}
 E(g)=\sum_{\rho} \dim(\rho) \dd_e(\rho) \delta_{R_\gamma}(\rho)
\int\frac{dm}{4\pi}\langle m|g^+|m\rangle^{2j^+} \langle \bar{m}|g^+|\bar{m}\rangle^{2j^-}.
\end{equation}

Now we want to compute
\begin{equation}
 e^\rho_{k}=\frac{1}{2k+1}{\rm Tr} I^\dagger(\rho,k)\tilde{E}_\rho
I(\rho,k)
\end{equation}
Note that $I(\rho,k)$ is given by its matrix elements in the coherent state basis,
thus if $\rho = (j^+,j^-)$ with $j^\pm=\frac{\gamma\pm 1}{2}\tilde{k}$ and
$j^+$, $j^-$, $k$ admissible we have that
\begin{equation}
\begin{split}
 &\left\langle I(\rho,k)|n\rangle^{2k}, |n^+\rangle^{2j_+}\otimes |n^-\rangle^{2j_-}\right\rangle\\
&=\frac{\sqrt{2k+1}}{C}\langle n,n^+\rangle^{2(k+j^+-j^-)}\langle
n,n^-\rangle^{2(k+j^--j^+)}
\epsilon(n^+,n^-)^{2(j^-+j^+-k)}
\end{split}
\end{equation}
with $\epsilon(\cdot,\cdot)$ being the invariant bilinear form and
\begin{equation}
 C^2=\frac{(j^++j^-+k)!(j^-+j^+-k)!(k+j^+-j^-)!(k+j^--j^+)!}{(2k)!(2j^+)!(2j^-)!}
\end{equation}
is derived in the book by Kauffman and Lins \cite{kauffman1994temperley}.

Since $I^\dagger(\rho,k)\tilde{E}_\rho I(\rho,k) $ is proportional to identity, we only need to find
\begin{equation}
 \left\langle \langle \coh|^{2k}, I^\dagger(\rho,k)\tilde{E}_\rho I(\rho,k)
|\coh\rangle^{2k}\right\rangle,
\end{equation}
for the case $\rho \in R_\gamma$. Then this is equal to
\begin{equation}
\begin{split}
  d_e\frac{2k+1}{C^2}\int \frac{dm}{4\pi}&
\langle \coh,m\rangle^{2(k+j^+-j^-)}\langle \coh,\bar{m}\rangle^{2(k+j^--j^+)}
\epsilon(m,\bar{m})^{2(j^-+j^+-k)}\\
&\langle m,\coh\rangle^{2(k+j^+-j^-)}
\langle \bar{m},\coh\rangle^{2(k+j^--j^+)}
\epsilon(m,\bar{m})^{2(j^-+j^+-k)}
\end{split}
\end{equation}
But we know
\begin{equation}
|\langle \coh,m\rangle|^2=\left(\cos\frac{\theta}{2}\right)^2,\quad
|\langle \coh,\bar{m}\rangle|^2=\left(\sin\frac{\theta}{2}\right)^2
\end{equation}
where $\theta$ is the angle between the direction of $m$ and the north pole. We can introduce polar
coordinates
\begin{equation}
\begin{split}
 e^\rho_k = \Big\langle \langle \coh|^{2k}&, I^\dagger(\rho,k)\tilde{E}_\rho I(\rho,k)
|\coh\rangle^{2k}\Big\rangle\\
&=d_e\frac{2k+1}{4\pi C^2}\int_0^{2\pi} d\phi\int_0^\pi d\theta \sin\theta
\left(\cos\frac{\theta}{2}\right)^{2(k+j^+-j^-)}\left(\sin\frac{\theta}{2}\right)^{2(k+j^--j^+)}\\
&=d_e\frac{2k+1}{4\pi C^2}2\pi 2\int_0^{\pi/2} d\eta 2\sin\eta\cos\eta
(\cos\eta)^{2(k+j^+-j^-)}(\sin\eta)^{2(k+j^--j^+)}\\
&= d_e\frac{2k+1}{C^2}\frac{
(k+j^+-j^-)!(k+j^--j^+)!}{(2k+1)!}
\end{split}
\end{equation}
We thus obtained
\begin{equation}
 e^\rho_{k}=d_e
\frac{(2j^+)!(2j^-)!}{(j^++j^-+k)!(j^-+j^+-k)!}
\end{equation}
Substituting $j^\pm=\frac{\gamma\pm 1}{2}\tilde{k}$ we have that for $\rho = (j^+,j^-)$ the basis coefficient is given by
\begin{equation}
e^\rho_{k}=d_e \frac{((\gamma+1)\tilde{k})!((\gamma-1)\tilde{k})!}
{(\gamma\tilde{k}+k)!(\gamma\tilde{k}-k)!}.
\end{equation}
For $\gamma>1$, $\tilde{k}$ is the minimal representation in
the decomposition and
\begin{equation}
 e^\rho_{\tilde{k}}=d_e
\end{equation}
Finally, making the conditions on $\rho$ explicit again we can write
\begin{itemize}
 \item FK$_{\gamma>1}$: $e^{\rho}_k = \dd_e(\rho)
\sum_{\tilde{k}}\delta\left(\left(\frac{\gamma+1}{2} \tilde{k},
\frac{\gamma-1}{2} \tilde{k}\right),\rho\right)
\frac{((\gamma+1)\tilde{k})!((\gamma-1)\tilde{k})!}
{(\gamma\tilde{k}+k)!(\gamma\tilde{k}-k)!}$
\end{itemize}

Thus the FK model also falls into the class of real holonomy spin foam models.

\subsubsection{Baratin-Oriti}

Let $\beta=\frac{\gamma-1}{\gamma+1}$. We will use the following $\gamma$ dependent transformation from $\SU(2)$ to $\SU(2)$:
{\begin{equation}
 SU(2)\ni \cos\theta+i\vec{n}\vec{\sigma}\sin\theta=u\rightarrow u^\beta=
\cos\theta_\beta+i\vec{n}_\beta\vec{\sigma}\sin\theta_\beta
\end{equation}
where the class angle $\theta_\beta$ and the unit vector $\vec{n}_\beta$ are
determined by conditions \cite{Baratin:2011hp}
\begin{equation}
 \sin\theta_\beta=|\beta|\sin\theta,\q \text{sign}(
\sin\cos\theta_\beta)=\text{sign}(\cos\theta),
\q \vec{n}_\beta=\text{sign}(\beta)\vec{n}\ .
\end{equation}
Let us notice that
\begin{equation}
 (u^\beta)^{-1}=(u^{-1})^\beta,\q (gug^{-1})^\beta=gu^\beta g^{-1}\ .
\end{equation}}

The fusion coefficients for the BO model are given in \cite{Baratin:2011hp} (eq.
37,58). From these one can derive
\begin{align}
  \tilde{E}^{j+j-}_{(m^+m^-)(\tilde{m}^+\tilde{m}^-)}&=
\int du\
D^{j^-}_{m^-n^-}(u^{-1})D^{j^+}_{m^+n^+}((u^\beta)^{-1})\nn\\
&\int d\tilde{u}\
D^{j^-}_{n^-\tilde{m}^-}(\tilde{u})D^{j^+}_{n^+\tilde{m}^+}(\tilde{u}^\beta)
\end{align}
Thus the $E$ function
\begin{equation}
 E(g^+,g^-)=\sum_{j^+j^-} \dim(j^+)\dim(j^-)
\tilde{E}^{j+j-}_{(m^+m^-)(\tilde{m}^+\tilde{m}^-)}
D^{j^+}_{\tilde{m}^+m^+}({g^+}^{-1})
D^{j^-}_{\tilde{m}^-m^-}({g^-}^{-1})
\end{equation}
is equal to
\begin{align}
 E(g^+,g^-)&=\sum_{j^+j^-} \dim(j^+)\dim(j^-)
\int dud\tilde{u}\
\chi_{j^-}(u^{-1}\tilde{u} {g^-}^{-1})
\chi_{j^+}((u^\beta)^{-1}\tilde{u}^\beta {g^+}^{-1})\\
&=\int dud\tilde{u}\
\delta(u^{-1}\tilde{u} {g^-}^{-1})
\delta((u^\beta)^{-1}\tilde{u}^\beta {g^+}^{-1})
\end{align}
We can simplify the equation by solving the first delta function for $\tilde{u}$ so that
$\tilde{u}=ug_-$
\begin{equation}
 E(g^+,g^-)=\int du\ \delta((u^\beta)^{-1}(ug^-)^\beta {g^+}^{-1})
\end{equation}
In the case of $\beta=1$ this $E$--function reduces to the one of the BC model.
{The BO model is also a real holonomy spin foam model.}

\section{The transfer operator for holonomy spin foams}\label{transfer}

So far we have discussed at length the structural aspects of holonomy spin foam models, their boundary Hilbert spaces and their gluings. This raises the question if, having the boundary Hilbert spaces at hand, we can define a Hamiltonian dynamics reflecting the one defined by the spin foam models.

The first step in deriving such a Hamiltonian dynamics from a given partition
function is to obtain transfer operators. In standard lattice theories, for
example lattice gauge Yang Mills theory, such transfer operators correspond to
finite time steps. To obtain the Hamiltonians as infinitesimal time evolution
generators for such systems one would have to take the limit of infinitesimal
time by scaling the coupling constants in time -- and space directions in a
certain way defined by the dynamics of the system \cite{ks,kr}

The issue of obtaining the Hamiltonians is more involved in gravitational systems, as the lattice constants and the time separation are rather encoded in the boundary states of the system. Furthermore the question of exact and broken diffeomorphism symmetry comes in \cite{bd1,b1,bd12}. Only in the case that exact diffeomorphism symmetry is preserved in the discretization, can we expect the appearance of Hamiltonian constraints in the canonical formulation \cite{b1,bd1}. If this holds also for the partition function the transfer operator is a product of projection operators from which the Hamiltonian and diffeomorphism constraints can be read off \cite{Noui:2004iy,Noui:2004ja, Steinhaus,Bahr:2011yc}. In this case no limiting procedure is necessary. If diffeomorphism constraints are broken, one can either attempt a limiting procedure involving the boundary states or alternatively attempt to obtain an improved model by coarse graining which then carries a notion of diffeomorphism symmetry \cite{bd1,bd2,Dittrich:2011zh,bd12,bd1205,Comp, tate}.

A third possibility is to adopt the view point that the dynamics is inherently discrete, a viewpoint which is for instance emphasized in the framework of consistent discretizations \cite{Gambini:2005sv,Gambini:2005vn}. In this case the transfer operator can only defined for finite time steps, and a limit cannot be taken (in general).

Here we will consider the finite time transfer operator and comment more on the issues of taking the limit to obtain the time evolution generators afterwards. We will consider a space time lattice with a regular slicing in time direction, i.e. each (thick) time slice  is of the form $\Gamma_s \times [0,1]$. The discussion can be generalized to some extend to an irregular lattice and a notion of local time evolution, see for instance \cite{hoehn1,hoehn2} for a discussion in classical Regge calculus.

The definition of the transfer operator requires a choice of slicing of the underlying lattice and initially we choose one
 which will make the transfer operator as similar as possible to the one encountered in lattice gauge theory \cite{ks,kr,smit,Bahr:2011yc}. As we will see such a slicing fits well to the universal boundary Hilbert space introduced in section \ref{sec-Boundaries}.  In this formulation the effective face weights introduced in \cite{Comp} and recalled in section \ref{sec-FaceToFace} will play a prominent role. From a simplicial geometry view point the faces are dual to the bones of the triangulation, which carry the curvature. Thus this slicing offers a new perspective on the Hamiltonian dynamics and the semiclassical limit: it does not concentrate on the vertex (i.e. simplex) amplitude but on the gluing of simplices around the bones, where the curvature and hence the essential dynamical information resides.

On the other hand this slicing is somewhat unusual in discrete gravity, where one often builds a transfer operator by gluing simplices to the hypersurface \cite{Alesci:2010gb,hoehn1,hoehn2}. In this case equal time hypersurfaces can be understood as (dual to) $(D-1)$ dimensional triangulations.
As we will see in the course of the discussion we can switch to a slicing more adapted to a simplicial viewpoint  by using the $\mu$ map between the projected spin network and universal Hilbert space introduces in section \ref{sec-TrimmedComplexes}. The $\mu$ map can then be understood to project onto the solutions of the simplicity constraints -- the (stripped) transfer operator will be sandwiched between such projectors.

\subsection{Transfer operator for general models}

Transfer operators can be defined if we have a discrete ``time'' direction in our 2-complex in the sense that the complex $\CC$ is the 2-skeleton of the complex $(\CC_s \times [0,1])^n$, with $\CC^n_s\times[1] = \CC^{n+1}_s\times[0]$. In that case we call the edges and faces in the various $\CC_s^n$ spatial, and the other edges and faces temporal.

We call the graph of horizontal edges $\Gamma^s$. Then the partition functions $\ZZ(\CC_s)$ and $\ZZ(\Gamma^s \times [0,1])$ act naturally on the space $\HH_{\mbox{UBS}}^{\Gamma^s}$, $\ZZ(\CC_s)$ simply by multiplication.

It follows directly from the gluing of in $\HH_{\mbox{UBS}}^{\Gamma^s}$ according to \eqref{eq-gluing} that the partition function of $\CC$ can then be written as

\be
\ZZ(\CC) = \left(\ZZ(\CC_s) \ZZ(\Gamma_s \times [0,1])\right)^n \ZZ(\CC_s)
\ee

This has the structure of a partition function written in terms of transfer operators. For the rest of this section we call $\ZZ(\CC_s) = W$, $\ZZ(\Gamma_s \times [0,1])= K$ and $\ZZ(\CC_s) \ZZ(\Gamma_s \times [0,1]) = T$, that is

\be
T = W K,
\ee

and

\be
\ZZ(\CC) = T^n W.
\ee

The operators $W$ and $K$ can be written very efficiently in terms of the effective face weights $\omega_f$ which we recalled in section \ref{sec-FaceToFace} in \eqref{eq-effectiveFaceWeights}. Recall that these are simply given by the amplitude of a face, as illustrated in Figure \ref{fig-Face}.

As the entire 1-skeleton of $\CC_s$ is in the boundary space the only integrations in $W$ are those involving the $g_{ef}$. These are exactly the integrations one performs to obtain the effective face weights $\omega_f$. The operator $W$ acts as a multiplication operator in the holonomy basis of the universal Hilbert space and is just given as a product of the effective face weights
\be
W[g_{ev}] = \prod_{f \in \CC_s} \omega_f(\{g\}^f),
\ee
where the notation $\{g\}^f$ indicates the set of all group elements belonging to the face $f$.

\begin{figure}[htbp]
 \centering
 \includegraphics[scale=.25]{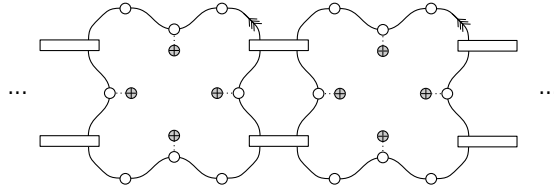}
 \caption{The composition of two effective face weights as it occurs in the definition of $K$.}
 \label{fig-K}
\end{figure}

For $K$ we have a similar simplification. Its structure is illustrated in figure \ref{fig-K}. In addition to the integrations over the $g_{ef}$ we have also to take into account the integrations over the $g_{ev}$ associated to the time like edges, which we will denote by $g_{ev}^t$. By ${g'}_{ev}^{s}$ and $g_{ev}^s$ we denote holonomy variables associated to edges in the spatial hypersurfaces at  two consecutive time steps. The integration  kernel of the operator $K$ in the holonomy representation is given by

\be
K[{g'}^{s}_{ev}, g^s_{ev}] = \int \left(\prod \dd g^t_{ev}\right) \prod_{f \in \Gamma_s \times [0,1]} \omega'_f(\{g\}^f)
\ee
where each effective face weight $\omega'_f(\{g\}^f)$ depends on four variables $g_{ev}^t$ associated to the two time like edges of the time like face, a set of variables ${g}_{ev}^s$ associated to the space like edges of this face shared with the graph $\Gamma_s \times [0]$ and a corresponding set ${g'}_{ev}^s$ shared with $\Gamma_s \times [1]$.

Each time like edge has one boundary vertex in $\Gamma_s\times [0]$ and one in $\Gamma_s\times [1]$, and each vertex in  $\Gamma_s\times [0]$ or $\Gamma_s\times [1]$ has only one temporal edge going out. Thus we can drop the $e$ in $g^t_{ve}$ and simply write $g^t_v$. 


Recall that the effective face weights $\omega_f$ naturally live in the universal boundary space of a segmented line, and thus have a gauge freedom that acts as $g_{ve} \rightarrow g_v g_{ve}$. Hence we can apply gauge transformations to the vertices in $\Gamma_s\times [0]$ and $\Gamma_s\times [1]$ such that all the group variables associated to the time like edges are equal to the identity. This allows us to write the integration over these group elements associated to the time like edges as gauge projectors on the vertices of the graph $\Gamma_s$
\be
K = P_G\, K_0\, P_G.
\ee
%
%
%
%
%
%
where
\be
P_G[{g'}^s_{ev}, {g}^s_{ev}] = \int_G \left(\prod_{v \in \Gamma_s} \dd g^t_v\right) (\prod_{e \in v} \delta({g'}^s_{ev} g^t_{v} {g}^s_{ve})),
\ee
is simply the projection on gauge invariant functions on the vertices of the graph.

$K_0$ is obtained by setting the group variables in $\omega_f$ or equivalently $\omega'_f$ associated to the time like edges equal to the identity element
\ba
K_0[{g'}^s_{ev}, {g}^s_{ev}]= \prod_e \omega_{f(e)} ( g_{v_1e},g_{ev_2},g'_{v_2e},g'_{ev_1})
\ea
where $v_1,v_2$ are the source and target vertex of the edge $e$ and $f(e)=e\times [0,1]$ is the time like face associated to the edge $e\in \Gamma_s$.

\begin{figure}[htbp]
 \centering
 \includegraphics[scale=.25]{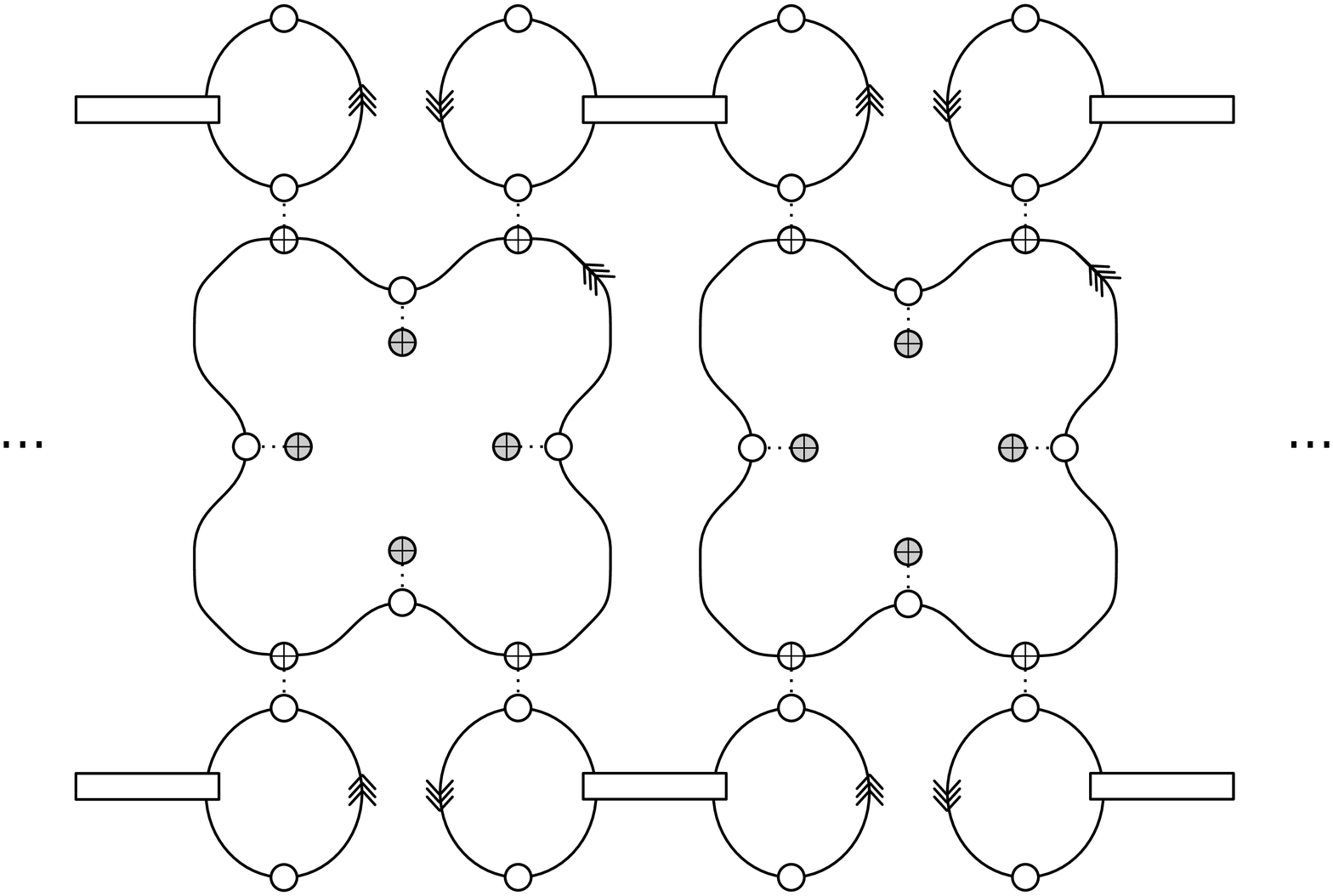}
 \caption{The decomposition of the effective face weights into reduced face weights and gauge projectors.}
 \label{fig-PKP}
\end{figure}

This structure is illustrated in figure \ref{fig-PKP}. A simple application of the top relation in figure \ref{fig-Relations} will return us to figure \ref{fig-K}.

The operator $K_0$ does not map from the universal boundary space to the universal boundary space as it does not produce a state with $G$ invariance at the vertices but acts in $L^2(G^{|\Gamma_{ev}|})$ instead. The projector $P_G$ then brings us back to the gauge invariant universal boundary space.




It will be convenient to write $K_0=\prod_e K_e$  as an operator in the usual
sense, in terms of left and right shift operators  $(L(g) \triangleright
f)(\cdot) = f(g^{-1}\cdot)$ and  $(R(g) \triangleright f)(\cdot) = f(\cdot
g)$ respectively. To this end we introduce a bra--ket notation such that we
write an element  $f(g_1,g_2)\in L^2(G\times G)$ as $f(g_1,g_2)=\langle
g_1,g_2|f\rangle$.

The operator $K_e$  can then be written as
\ba\label{13}
K_e&=&\int dg_A dg_B \prod_{i=1}^4 \left(dg_i \,E(g_i)\right)
\,\omega_f(g_Bg_A^{-1})\, L_1(g_4)R_1(g_1g_A) L_2(g_B g_3) R_2(g_2). \q\q
\ea
where the left and right shift operators act as
\ba\label{12}
L_1(g_a) R_1(g_b) L_2(g_c) R_2(g_d) \,|g_{1},g_{2}\rangle &=& |g_a^{-1}
g_{1} g_b ,\, g_c^{-1} g_{2} g_d\rangle  \q . \q\q
\ea

In summary the transfer operator is given by
\ba\label{sumT}
T=W\cdot K = \left(\prod_{f \in \CC_s} \omega_f \right) \left( \prod_{v \in \CC_s} P_{G,v}  \right) \left(\prod_{e \in \CC_s} K_e\right)\left( \prod_{v \in \CC_s} P_{G,v} \right)
\ea
where $W$ acts as a multiplication operator in the holonomy basis and factorizes over the faces, $P_{G,v}$ is the projector on the $G$--gauge invariant subspace at the vertex $v$ and we defined the action of $K_e$ in equation (\ref{13}).

\subsection{Transfer operator in the spin network basis and simplicity constraints}

The operator $K_e$ is simplest in the spin basis. As $K_e$ just acts on one edge it is sufficient to
consider the one--edge Hilbert space $L^2(G\times G)$. A basis for this space would be given by $|\rho_1,i_1,j_1;\rho_2,i_2,j_2\rangle$, where $i_1,j_1,i_2,j_2$ are magnetic indices in the representations $\rho_1,\rho_2$ respectively.

However, to compactify notation it is useful to introduce a basis adapted  to the $\H$ group in a
given $\rho$ representation. The basis
we introduce is of Gelfand-Tsetlin type \cite{GT}. It is labelled by
\begin{equation}
 j\leftrightarrow \{k,d,m\}, \quad |\{k,d,m\}\rangle =I(\rho,k)^d |m\rangle
\end{equation}
where $k$ is the label of the $\H$ representation, $d$ is the multiplicity index (in the
multiplicity free case it will be omitted), and $|m\rangle$ is a basis in the $k$ representation. Thus we replace the (four) magnetic indices of the $\rho$--representations in $L^2(G\times G)$ by four indices $j_i\leftrightarrow \{k_i,d_i,m_i\}$.

{The basis in the space of $\H$ invariant functions is thus spanned by
\begin{equation}
 |\rho_1,\{k_1,m_1\};k;\rho_2,\{k_2,m_2\}\rangle=\frac{1}{\sqrt{\dim(k)}}\sum_{m}
 |\rho_1,\{k_1,m_1\},\{k,m\};\rho_2,\{k_2,m_2\},\{k,m\}\rangle
\end{equation}
the same functions as introduced in \eqref{snbasis1}.}

As follows from \eqref{13} and picture \ref{fig-PKP} the operator $K_e$can be decomposed into the following more elementary components
\begin{equation}
 K_e= K^s_1 K^s K_3^s\ K^t_2 K^t_4\ ,
\end{equation}
where
\begin{equation}
K^s=\int dg_Adg_B\ \omega_f(g_Bg_A^{-1}) R_1(g_A)L_2(g_B)=
\int dg \omega_f(g) R_1(g)\ \int dg'R_1(g')L_2(g')
\end{equation}
and
\begin{align}
&K^s_1=\int dg_1\ E(g_1) R_1(g_1)
&K^t_2=\int dg_2\ E(g_2) R_2(g_2)\\
&K^s_3=\int dg_3\ E(g_3) L_2(g_3)
&K^t_4=\int dg_4\ E(g_4) L_1(g_4)\,.
\end{align}
Let us notice that $K^t_2$ and $K^t_4$ commute with the rest of the operators and $K^s_1 K^s K_3^s$
commutes with $P_G$.

We also have
\begin{equation}
K^s=\sum_\rho  \tilde{\omega}^{\rho} P_\rho
\end{equation}
where $P_\rho$ is the projection onto the subspace in $L^2(G\times G)$ spanned by the following orthonormal basis labelled by
$j_1,j_2$
\begin{equation}
 \frac{1}{\sqrt{\dim(\rho)}}\sum_{j}|\rho, j_1, j; \rho, j,j_2\rangle \q .
\end{equation}


The operators $K^{s/t}_n$ act on the basis $|\rho_1, i_1, j_1; \rho_2, i_2,j_2\rangle$ as follows.
The operator $K^s_1$ changes only the index $j_1=\{k,d,m\}$ by multiplication of the matrix
\begin{equation}
 e^k_{\rho_1\ \tilde{d} d}\delta_{\tilde{m}m}\delta_{\tilde{k}k} \q .
\end{equation}
Similarly the operators $K^t_2, K^s_3, K^t_4$ act only on the indices $j_2,i_2, i_1$ respectively.

In the multiplicity free case we have a straightforward eigenfunction expansion of both $K^t_2K^t_4$ and
$K_1^sK^sK^s_3$.
The eigenvectors with non--vanishing eigenvalues for $K_1^sK^sK^s_3$ are given by
\begin{equation}\label{sepp1}
 |\rho,\{k_1,m_1\};\{k_2,m_2\}\rangle=\frac{1}{\sqrt{\sum_k \dim(k) (e^k_\rho)^2}}
\sum_{k,m} e^k_\rho|\rho,\{k_1,m_1\},\{k,m\};\rho,\{k,m\},\{k_2,
 m_2\}\rangle
\end{equation}
with corresponding eigenvalues
\begin{equation}
 K_1^sK^sK^s_3|\rho,\{k_1,m_1\};\{k_2,m_2\}\rangle=\frac{\tilde{\omega}^\rho}{
\dim(\rho) } \left(\sum_k \dim(k)
(e^k_\rho)^2\right)
|\rho,\{k_1,m_1\};\{
k_2,m_2\}\rangle \, .
\end{equation}
These are also eigenvectors for $K_2^tK_4^t$ with eigenvalues
\begin{equation}
 e^{k_1}_\rho e^{k_2}_\rho \q .
\end{equation}
In summary the eigenvectors with a priori non--vanishing eigenvalues for  $K_e$ are given by (\ref{sepp1}) with eigenvalues
\begin{equation}\label{sepp2}
 e^{k_1}_\rho e^{k_2}_\rho\ \frac{\tilde{\omega}^\rho}{\dim(\rho)}\left(\sum_k
\dim(k)(e^k_\rho)^2\right) \q .
\end{equation}
Thus the eigenvalues are independent  of the labels $m_1,m_2$ and come with a multiplicity $\dim(k_1)\dim(k_2)$. Note that $K_e$ vanishes on states $|\rho_1,i_1,j_1;i_2,j_2\rangle$ with $\rho_1\neq \rho_2$ as well as on states with $\rho=\rho_1=\rho_2$ but orthogonal to (\ref{sepp1}).

\bline

In the gravitational models the $E$ functions and therefore the $e^\rho_k$   impose the simplicity constraints. Hence we can say the same of $K_e$ -- it maps onto a subspace of the universal boundary Hilbert space on which the (primary) simplicity constraints hold in some form\footnote{As the discrete form of the primary simplicity constraints do not even commute weakly one has the choice to impose them strongly, i.e.  as operator equations, as in the BC model \cite{Barrett:1999qw} or in a certain weak form as in the EPRL model \cite{Engle:2007qf}.}.
As the $K_e$ map to a subspace which can be interpreted as solutions to the primary simplicity constraints let us also consider the question whether $W$, or some suitable subset of holonomy operators, leaves this subspace of the universal boundary Hilbert space invariant.

$W$ is a multiplication operator that factorizes over the spatial plaquettes. The contribution from a given plaquette is of the form
\begin{equation}\label{23}
 w'_{f}=  \sum_{\rho}  \dim(\rho)\,\tilde \omega^\rho \,\, \prod_{e\in f_s}
D_{\rho\ {\{k_{ev}', d_{ev}', m_{ev}'\}}}^{\{k_{ev}, d_{ev}, m_{ev}\}}(g_{ev}^{-1})
e^{k_{ev}}_{\rho, d_{ev}d_{ev'}} \delta_{k_{ev}'k_{ev'}'}\delta_{m_{ev}'m_{ev'}'}
D_{\rho\ {\{k_{ev'}, d_{ev'}, m_{ev'}\}}}^{\{k_{ev'}', d_{ev'}', m_{ev'}'\}}(g_{ev'})
\end{equation}
where the edge $e$ joins $v$ with $v'$. From now on we will consider only the multiplicity free case, thus
omit the index $d$.

These are contractions between $\delta_{\rho_1 \rho}\delta_{\rho_2\rho} e^\rho_k$ and  basic holonomy operators
\begin{equation}\label{25}
 \psi_{\rho_1,k_1,m_1\,;k\,;\,\rho_2,  k_2,m_2}= \sum_m D_{\rho\ {\{k, m\}}}^{\{k_1,
 m_1\}}(g_{ev_1}^{-1})
D_{\rho\ {\{k_2,, m_2\}}}^{\{k, m\}}(g_{ev_2}) \q .
\end{equation}
{Note that $\psi_{\rho_1,k_1,m_1\,;k\,;\,\rho_2, 
k_2,m_2}$ acting on the constant function creates the states
$$|\rho,\{k_1,m_1\};k;\rho_2,\{k_2,m_2\}\rangle.$$}


Thus the action of $W$ involves the multiplication of holonomy operators of the form (\ref{25}). We will therefore consider the product of two such holonomy operators. This is a straightforward calculation, in which one first rewrites
\ba\label{26}
D_{\rho\ j}^{i}(g)\,D_{\rho'\ j'}^{i'}(g)=\sum_{\rho'',i'',j''} \bar
C^{\rho\rho'\rho''\ ii'}_{\phantom{\rho\rho'\rho''}i''}\,C^{\rho\rho'\rho''\
j''}_{\phantom{\rho\rho'\rho''}jj'} \,D_{\rho''\ j''}^{i''}(g)
\ea
with $C^{\rho\rho'\rho''}_{jj'j''} $ the Clebsch Gordan coefficients of $G$.
Here we assume that $G$ is multiplicity free, i.e. there is maximally one copy of a given irrep in
the tensor product of two irreps.  We will assume the same property to hold for the subgroup  $\H$,
furthermore we already assumed that there is maximally one copy of a given $\H$ representation$k$
in a given $G$ representation.

In this case the Clebsch Gordan coefficient for $G$ contracted with the
maps $I(\rho,k)$ reduce to the Clebsch Gordan coefficients of $\H$ which we can write in the
Gelfand-Tsetlin-like basis as
\begin{equation}\label{27}
C^{\rho\rho'\rho''\phantom{\{k,m\},\{k',m'\}}\{k''.m''\}}_{\phantom{\rho\rho'\rho''}
\{k,m\},\{k',m'\}}= C^{kk'k''\phantom{mm'}m''}_{\phantom{kk'k''}mm'} \q .
\end{equation}

Finally the summation over the index $m$ in the holonomy operators (\ref{25}) leads to the following contraction of Clebsch Gordan coefficients
\ba\label{28}
\sum_{m,m'} C^{kk'k''}_{mm'm''} \bar C^{kk'k''}_{mm' \tilde m''} =\theta(k,k',k'')\,\delta_{m'',\tilde m''}
\ea
where $\theta(k,k',k'')=1$ if $k,k',k''$ couple to the trivial representation and is vanishing otherwise.

The product of two holonomy operators of the form (\ref{25}) is therefore given by
\ba\label{29}
&&\psi_{\rho_1,k_1,m_1\,;k\,;\,\rho_2,  k_2,m_2}\,\times \,\psi_{\rho'_1,k'_1,m'_1\,;k'\,;\,\rho'_2,  k'_2,m'_2} \nn\\
&=&
\sum_{\rho''_1,\rho''_2,k'',k_1'',k_2''}     \bar C^{k_1k'_1k''_1}_{m_1m_1'm_1''}     \,    \theta(k,k',k'')   \,   C^{k_2k_2'k_2''}_{m_2m_2'm_2''}   \,\,   \psi_{\rho''_1,k''_1,m''_1\,;k''\,;\,\rho''_2,  k''_2,m''_2}  \q .
\ea
Here we sum over all repeated magnetic indices. The result is again a linear combination of basis states (\ref{25}).

From this expression (\ref{29}) we notice the following: (a) Even if initially the representations satisfy $\rho_1=\rho_2$ and $\rho_1'=\rho_2'$ this will in general not hold for the basis states appearing on the right hand side of (\ref{29}). (b) Consider the case that we multiply two basic holonomies (\ref{25}) satisfying $\rho=\rho_1=\rho_2$ and $\rho'=\rho'_1=\rho'_2$ which have been contracted with $e^\rho_k$ and $e^{\rho'}_{k'}$ in the $k$ and $k'$ index respectively. The basis states in the product holonomy are then contracted with
\ba\label{30}
\sum_{k,k'} e^\rho_k \,e^{\rho'}_{k'} \,\,\theta(k,k',k'')
\ea
in the $k''$ index. In general the product of holonomy operators of the form (\ref{25}) will not generate a proper subspace. An exception is the Barrett Crane model, in which the $E$--function has an enhanced symmetry that allows a restrictions to spin network states with $k=0$. In this case $\theta(0,0,k'')\neq 0$ indeed leads to the condition $k''=0$.

The fact that the holonomy operators (\ref{25}) do not lead in general to a
proper subspace might not be a surprise to the expert as the secondary simplicity
constraints, which are conditions on the holonomies
\cite{zapata,alexandrovr,ryan1,ryan2} are usually not imposed in spin foam
models. Indeed the hope is that the imposition of the primary simplicity
constraints on two consecutive time slices leads to the automatic imposition of
the secondary constraints, whose function in a canonical formulation is to
ensure that the primary simplicity constraints are preserved under time
evolution. Later on we will redefine the transfer operator to make this notion
more explicit. Here we just note that one possibility is to consider
$T'=K_0^{\frac{1}{2}}P_G W  P_G K_0^{\frac{1}{2}}$ as we can interpret $K_0$ to
impose the simplicity constraints.

\bline

The  operator $W$ naturally factorizes over plaquettes and we can expect that it  leads to the curvature term $F$ in the gravitational Hamiltonian constraints, which are of the form $FEE$ with $E$ representing flux (infinitesimal shift) operators. On the other hand we can also seek an expression which factorizes over the vertices of $\Gamma_s$.  Such a form brings as back to the usual vertex amplitude representation of spin foams and shows the consistency of the procedure.

For this calculation the gauge invariance at the vertices of $\Gamma_s$ is essential, hence we
choose the basic states $|\rho_{ev},\eta_v,k_e\rangle$ introduced in (\ref{snbasis3}) which, using the GT basis, are defined by
\begin{align}
|\, \rho_{ev},\eta_v,k_e\rangle =&
\frac{\prod_{(ev)}\sqrt{\dim(\rho)_{ev}}}{\prod_{e}\sqrt{\dim(k_e)}}\nn\\
&\prod_{v} \eta_{v,j_{ev},\dots} \prod_{e}
D_{\rho_{ve}}(g_{ve})_{j_{ve}\{k_e,m_e\}}
D_{\rho_{ev'}}(g_{ev'})_{\{k_e,m_e\}j_{ev'}}\q .
\end{align}
Let us note that $W$ preserves the gauge invariant ${\mathcal H}_{UBS}$ (with respect to the gauge action at the vertices of the underlying graph).

We want to compute the matrix elements
\ba\label{au1}
\langle\rho'_{ev},\eta'_v, k'_e| W |\rho_{ev},\eta_v, k_e\rangle = \langle\rho'_{ev},\eta'_v, k'_e| P_GWP_G |\rho_{ev},\eta_v, k_e\rangle \q .
\ea

$W$ is a multiplication operator in the holonomy basis, hence to compute the matrix elements we introduce two resolutions of unity into the matrix elements (\ref{au1}). These resolutions of unity lead to an integration over the group elements $g_{v_1e}$ and $g_{ev_2}$. The holonomy operator associated to a given edge $e$ is then given by
\begin{equation}
 (W_e)_{\{i_{fe}i_{f(e+1)}\}}(\{\rho_f\})  =\ \prod_{f \ni e}  \sum_{k_{fe}, m_{fe}}
D_{\rho_f}(g_{v_1e})_{i_{ve},\{ k_{fe},m_{fe}\}} e^{\rho_f}_{k_{fe}}
D_{\rho_f}(g_{ev_2})_{\{ k_{fe},m_{fe}\}, i_{f(e+1)}}
\end{equation}
where the contribution $(W_e)$ come with  magnetic indices $\{i_{fe}i_{f(e+1)}\}$ that are
contracted between the different edges of a face.
(Relative orientation of the edges and the face is unimportant if the model is real, otherwise we assume here that these orientations agree.) The $W_e$ are contracted and then
summed over the $\rho_f$ (multiplied with $\tilde \omega^{\rho_f} \dim(\rho_f) $) to obtain the
full operator $W$.

\begin{figure}[htbp]
 \centering
 \includegraphics[scale=.6]{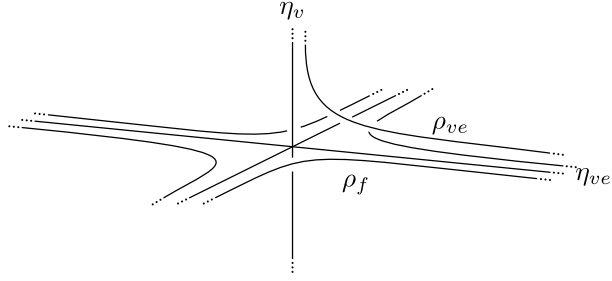}
 \caption{The labelling of the faces and edges around a spatial vertex.}
 \label{WinS}
\end{figure}

The computation then proceeds in the following steps which are completely analogous to the construction of vertex amplitudes in spin foam models, see for instance \cite{Bahr:2011yc,Perez:2012wv}. \\
(a) The integration over the group elements  $g_{v_1e}$ and $g_{ev_2}$ leads to the Haar projector
\ba
P_{i_1j_1\cdots i_n j_n}(\rho_1,\ldots,\rho_n)=\int dg \,D_{\rho_1}(g)_{i_1j_1}\cdots D_{\rho_n}(g)_{i_nj_n}
\ea
 on each half edge.
For instance for the first half edge the projector is on the invariant subspace in the tensor product
\be
V_{\rho_{v_1e}}\otimes V_{\bar\rho'_{v_1e}} \otimes { \bigotimes }_{f\ni e} V_{\rho
_f}  \q .
\ee
(b) The Haar projectors on each half edge can be split into a sum over a basis of orthonormal invariant vectors or intertwiners $\eta$ of the corresponding representation space.
\ba\label{au2}
P_{i_1j_1\cdots i_n j_n}(\rho_1,\ldots,\rho_n) =\sum_\eta  |\eta\rangle_{i_1\cdots i_n}  \langle \eta|_{j_1\cdots j_n} \q .
\ea
(c) The magnetic indices of the invariant vectors on the left in (\ref{au2}) associated to half edges $v_1 e$ and of the invariant vectors on the  right in (\ref{au2}) associated to half edges $ev_2$  contract now at the vertices of the graph $\Gamma_s$. Thus all the magnetic indices of $(W_e)_{\{i_{fe}i_{f(e+1)}\}} $ contract among each other. The magnetic indices $i_{v_1e},j_{ev_2}, i'_{v_1e},j'_{ev_2}$ associated to the representation matrices of our spin networks are contracted with the intertwiners $\eta_v,\eta'_v$  of these spin networks.\\
(d) The resulting amplitude is the vertex amplitude  $A^{BF}_v$ for $G$--BF theory. Here a vertex of $\Gamma_s$ is to be understood as a vertex in the following 2--complex: The edges $e$ of $\Gamma_s$ are `horizontal' edges in this two--complex and labelled with interwiners $\eta_{ve}$, which appeared in the expansion of the Haar projectors. Additionally we have the spatial faces, on which $W$ is defined and which are labelled by $\rho_f$. There are additional `vertical' edges and faces, which carry the labels of the spin networks between which we compute the matrix elements. For each $v\in \Gamma_s$ we have an edge pointing down and labelled with $\eta'_v$ and an edge pointing up labelled by $\eta_v$. There are also two vertical (half) faces attached to each (half) edge $(ve) \in \Gamma_s$ which are labelled by $\rho_{ve}$ and $\rho'_{ve}$. The orientation of these faces is such that these agree with the orientation of $e$ for the `up'  faces and are opposite with respect to the orientation of $e$ for the 'down' faces. This 2--complex around a vertex is depicted in figure \ref{WinS}

Thus $A^{BF}_v$ depends on all the algebraic data attached to (half) edges and faces adjacent to the vertex $v$ in this 2--complex.  This includes the intertwiners $\eta_{ve},\eta_v,\eta'_v$ which involve the representation spaces $\rho_f$ for faces sharing $f$ as well as $\rho_{ve},\rho'_{ve}$ for edges in $\Gamma_s$ sharing $v$. The vertex $BF$ amplitude is then defined by the contraction of the invariant vectors in the corresponding representation spaces
\be
A^{BF}_v (\eta_{ev}, \eta_v, \eta_v') = \tr_{\rho_{ev}} \tr_{\rho'_{ev}} \tr_{\rho_f \otimes \rho_{f'} \otimes \dots}  \,\,\eta_v \eta_{v'} \prod_{ev} \eta_{ev}.
\ee

\noindent
(e) We are left with half of the invariant vectors arising from the Haar projectors associated to the edges of $\Gamma_s$. These are contracted with the $I(\rho_f,k_{fe})_{j_{fe}m_{fe}} e^{\rho_f}_{k_{fe}} I^\dagger(\rho_{f},k_{fe})_{m_{fe}  i'_{fe}}$ part of the operators $W_e$ as well as with the $\sum_{m_e} I(\rho_{v_1e},k_{e})_{i_{v_1e}m_{e}} I^\dagger(\rho_{ev_2},k_{e})_{m_{e}i_{ev_2}}$ part of the spin network state $|\, \rho_{ev},\eta_v,k_e\rangle $ and the corresponding primed counterpart. Thus to each edge in $\Gamma_s$ we associate the amplitude
\ba
\!\!\! \!\!\!\!&&(P_e)^{\{\rho_{f}\}_{f\ni e} ,\rho_{v_1e},\rho'_{v_1e},\rho_{ev_2},\rho'_{ev_2}}_{\{k_{ef}\}_{f\ni e},k_e,k'_e}(\eta_{v_1e},\eta_{ev_2}) = \nn\\ &&
\!\!\!\!\!\!\!\langle\eta_{v_1e}| \bigg ( I(\rho_{ev_2},k_{e}) I(\rho_{v_1e},k_{e})^\dagger\otimes I(\rho'_{v_1e},k'_{e}) I(\rho'_{ev_2},k'_{e})^\dagger  \bigotimes_{f \ni e} e^{\rho_{f}}_{k_{ef}} I(\rho_{f},k_{ef}) I(\rho_{f},k_{ef})^\dagger \bigg )|\eta_{ev_2}\rangle . \nn\\
\ea

\noindent
(f) Finally the matrix elements of $W$ are given by
\ba\label{au3}
&&\langle\rho'_{ev},\eta'_v, k'_e| W |\rho_{ev},\eta_v, k_e\rangle = \nn\\
 &&\sum_{\eta_{ev}, \rho_{f}, k_{ef}}{\cal N} \, \prod_v A^{BF}_v (\eta_{ev}, \eta_v, \eta_v')  \prod_e (P_e)^{\{\rho_{f}\}_{f\ni e} ,\rho_{v_1e},\rho'_{v_1e},\rho_{ev_2},\rho'_{ev_2}}_{\{k_{ef}\}_{f\ni e},k_e,k'_e}(\eta_{v_1e},\eta_{ev_2})
\ea
where ${\cal N}$ collects all the dimension factors and face amplitudes
\ba
{\cal N}= \prod_f \dim(\rho_f)  \,\tilde{\omega}^{\rho_f}  \,\,\prod_{ev}   \sqrt{ \frac{\dim(\rho_{ev})} {\sqrt{\dim(k_e)}   }}   \sqrt{ \frac{\dim(\rho'_{ev})} {\sqrt{\dim(k'_e)}   }}  \q .
\ea

Thus the matrix elements of $W$ provide us almost with the full spin foam amplitude for an equal time slice. What is missing to obtain the full amplitude are the insertions of $e^\rho_k$ for the edge face pairs consisting of time like faces and space like edges as well as the edge amplitudes $P_e$ for the time like edges. These are indeed provided by the operators $K_e$.

%
%

%
%
%
%

\subsection{The transfer operator on the projected spin network Hilbert space}

So far we discussed the transfer operator in the unsymmetric form $T=W\cdot K$. In lattice gauge theory one often chooses rather a symmetric form $T_{LG} =W^{1/2}KW^{1/2}$. In the case of spin foams $W$ might not be a positive operator. Indeed for the gravitational spin foams it is rather easy to construct a square root of $K_e$ so that we can consider $T'=\prod_e K_e^{1/2} P_G WP_G \prod K_e^{1/2}$. We have seen that $K_e$ is almost a projection operator and have given the eigenvectors and eigenvalues (\ref{sepp1},\ref{sepp2}). Indeed for the EPRL model the non-null eigenvectors (\ref{sepp1}) reduce to
\be\label{veprl}
v^{EPRL}_{k,m_1,m_2}:=\left|\rho=\left(\frac{1+\gamma}{2}k, \frac{|1-\gamma|}{2}k
\right),\{k,m_1\};\{k,m_2\} \right\rangle
\ee
so that the index structure is the same as for (non-gauge invariant) $\H$ spin network functions,
i.e. the spin network basis of the standard loop quantum gravity Hilbert space. By formally
identifying the eigenvectors with this spin network basis we can define the transfer operator on the
LQG Hilbert space. In principle this applies also to the Barrett Crane model, the non--null
eigenvectors are however just labelled by one $\H$ representation label
\be\label{vbc}
v^{BC}_k:=|\rho=(k,k),\{0,0\},\{0,0\}\rangle
\ee
whose functionality is in addition quite different from the EPRL model.

A more elegant and geometric method is to use the $\mu_{\Gamma_S}$ map introduced in section \ref{sec-TrimmedComplexes}, which maps from the Hilbert space of projected spin network functions to the universal boundary Hilbert space. (As in this section we will therefore assume  that the face weights are given by delta functions $\omega_f=\delta_G$, however as mentioned there, this can be generalized.) The $\mu$ map was defined such that the effect of the time like plaquettes, i.e. $K$, can be written as
\ba
K=\mu_{\Gamma_s} \mu^\dagger_{\Gamma_s}.
\ea
In the following we will suppress the index $\Gamma_s$.

%




Thus we can define the transfer operator in the projected spin network space by
\be
\tilde\ZZ(\CC_t) = \left(\mu^\dagger W \mu \right)^n \q \text{i.e.}\q T_{PSN}=\mu^\dagger W \mu \q .
\ee

Remember that the integration kernel of $\mu$ is given by
\ba
\langle g_{v_1e},g_{ev_2}|\mu|\tilde  g_{v_1ev_2}\rangle
\!\!\!\!\!&& = \int \prod_{e \in \Gamma_s} dg_{v_1 e v_2} \prod_{(ve) \in \Gamma_s} dg^e_{vt} \, \prod_{v \in \Gamma_s} d g_{vt} \nn\\
 &&\prod_{e \in \Gamma_s} \delta(g_{v_1 e} g_{v_1ev_2} g_{ev_2} g_{v_2 t} g_{v_2 t}^e \tilde g_{v_2 e v_1} g_{t v_1}^e g_{t v_1}) E(g_{v_1ev_2}) F(g_{v_2 t}^e) F(g_{t v_1}^e )\q\q\q
\ea
where $F$ satisfies  $\int E(g) f(g) dg =\int
F(g)F(g^{-1}g')f(g')dgdg'$. $F$ can be expanded in the same way as $E$ and for
the coefficients we have $e^\rho_k=f^\rho_k f^\rho_k$. In the following we will
assume that $E$ is a projector, i.e. $f^\rho_k=e^\rho_k$. Here $t$ indicates a
time like edge, i.e. $g_{vt}$ is a group element associated to a time like
edge starting at the vertex $v$, $g_{tv'}$ a group element along a time like
edge ending in $v'$.

In the same way as for the operator $K$ we can extract a gauge invariant projector from $\mu$, i.e.
\ba
\mu=P_G \prod_e \mu_e
\ea
where $\mu_e$ is given by
\ba
\langle g_{v_1e},g_{ev_2}|\mu_e|\tilde  g_{v_1ev_2}\rangle =
\!\!\!\!\!&&\int  dg_{v_1 e v_2}  dg^e_{v_1t}  dg^e_{v_2 t}  \nn\\
 && \delta(g_{v_1 e} g_{v_1ev_2} g_{ev_2} g_{v_2 t}^e \tilde g_{v_2 e v_1} g_{t v_1}^e ) E(g_{v_1ev_2}) F(g_{v_2 t}^e) F(g_{t v_1}^e )\q .\q\q
\ea
In a spin network basis of the universal boundary Hilbert space  and the projected spin network space respectively we obtain
\begin{align}
&\langle\rho_1,\{k_1,m_1\}\,;k\,;\,\rho_2,  \{k_2,m_2\}|\mu_e|
\rho',\{k_1',m_1'\},\{k_2',m_2'\}\rangle=\nn\\
&\q\q\q\q \q\q\q\q\q\q\q\q\q\q\q\q\sqrt{\frac{\dim(k)}{\dim(\rho')}} \delta_{\rho_1\rho'}\delta_{\rho_2\rho'}\,
\,e^{\rho'}_{k}     \,   f^{\rho'}_{k_1}\delta_{k_1 k_1'} \,\, f^{\rho'}_{k_2} \delta_{k_2k_2'}  \q
.
\end{align}
 where we also
used a Gelfand-Tsetlin basis for the non--gauge invariant projected spin networks.

Thus the image of $\mu_e$ is spanned by the non-null eigenvectors for $K_e$ discussed previously and given in (\ref{veprl}) and (\ref{vbc}) for the EPRL and BC model respectively. The co--kernel of $\mu_e$ or image of $\mu_e^\dagger$  in the projected spin network space is labelled by the same indices, i.e.
\begin{align}
\tilde v^{BC}_{k}&=|(k,k),\{0,0\},\{0,0\}\rangle \nn\\
\tilde v^{EPRL}_{k,m_1,m_2}&=\left|\left(\frac{1+\gamma}{2}k, \frac{|1-\gamma|}{2}k
\right),\{k,m_1\},\{k,m_2\}\right\rangle \q .
\end{align}

Thus we can for instance for the EPRL model formally understand the transfer operator as an operator on the LQG Hilbert space. This might however not be very useful, if one wants to understand the structure of the transfer operator in terms of holonomy and flux operators. In particular the holonomy operators appearing in the transfer operator are $G$ holonomies and thus act on either the universal boundary Hilbert space or the projected spin network space, both of which are Hilbert space over copies of the group $G$.

\subsection{Example: The BF model}

Let us first consider the BF model, see also \cite{Bahr:2011yc} for a discussion of the corresponding transfer operator in the context of standard lattice gauge theory. $BF$ theory will be the only case where the transfer operator will be actually a  projector.

In the case of the BF model we have $e^\rho_k=\theta(\rho,k)$, where $\theta(\rho,k)=1$ if $k$
appears in the reduction of $\rho$ over $\H$ and $\theta(\rho,k)=0$ otherwise.  Also, as the face
weights are given by delta functions on the group we have $\tilde\omega^\rho=1$.

$K_e$ projects onto states
\ba\label{31}
v_{\rho,k_1,m_1;k_2,m_2} &=&\sum_k \sqrt{\dim(k)}\theta(\rho,k) \,
|\rho,\{k_1,m_1\}\,;k\,;\,\rho,
\{k_2,m_2\}\rangle \nn\\
&=&
\sum_{i_1,j_1,i_2,j_2,m} I^\dagger(\rho,k_1)_{m_1i_1}\delta_{j_1i_2}  I(\rho,k_2)_{j_2m_2} \,|\rho,i_1,j_1;\rho,i_2,j_2\rangle
\ea
which are just linear combinations of spin network states that have been subdivided with a trivial two--valent edge. The eigenvalues of $K_e$ are given by
\ba\label{32}
\lambda_{\rho,k_1,m_1;k_2,m_2}=  \frac{1}{\dim(\rho)}  \,\, 
\theta(\rho,k_1) \,\, \theta(\rho,k_2) \sum_k
\dim(k)  \, \theta(\rho,k)^2 \,\,=\,\,  \theta(\rho,k_1) \,\, \theta(\rho,k_2)  \, ,
\ea
i.e. are equal to one as long as the indices in $v_{\rho,k_1,m_1;k_2,m_2}$
define a non-vanishing vector.  Thus $K_e$ is a proper projection operator and
acts as identity on states that can be embedded into the standard lattice gauge
theory Hilbert space for $G$.

The operator $W$ is a multiplication operator in the holonomy basis given by the effective face weights associated to the spatial plaquettes. As $E(g)=\delta(g)$ the effective face weights are also given by delta functions evaluated on the holonomy around the plaquette
\ba\label{33}
\omega'_f(g_{v_1e_1}g_{e_1v_2},g_{v_2e_2} \cdots)\;=\;\delta( g_{v_1e_1}g_{e_1v_2} g_{v_2e_2} \cdots) \q .
\ea
Hence $W$ is a projection\footnote{As it involves the delta function it is not a
proper projection operator in the case of Lie groups. A mathematical clean
description can be obtained by interpreting $W$ as rigging map which maps to the
dual of some dense subspace of the Hilbert space, see for instance
\cite{Noui:2004iy}.} onto the states satisfying the flatness conditions.

Hence the transfer operator $T$ is a projector implementing the Gauss constraints, which impose gauge invariance, as well as the flatness constraints on the plaquettes.

\subsection{Example: The BC model}

Let us also discuss the Barrett Crane model as here an enhanced symmetry of the $E$ function allows to reduce the boundary Hilbert spaces.

For the Barrett Crane model the $e^\rho_k$ coefficients factorize $e^\rho_k=\delta_{k,0} \sum_{k'}\delta_{\rho,(k',k')}$. The operator $K_e$ projects onto states
\ba\label{34}
v_{k'}= |(k',k'),\{0,0\};0;(k',k'),\{0,0\}\rangle
\ea
and the corresponding eigenvalue is given by
$
\lambda_{k'}=\frac{1}{(\dim(k'))^2}
$.

More in general we can see that the holonomy operators (\ref{25}) with $k=0$, that is  $\psi_{\rho_1,k_1,m_1\,;0\,;\,\rho_2,  k_2,m_2}$  generate a closed subspace.
Hence the dynamics only involves this subspace of the universal boundary Hilbert space. Indeed the
integrand of the partition function, that is the effective face weights do show an enhanced symmetry
in the case of the Barrett Crane model: The $E$-function is given as
$E(g^L,g^R)=\delta_{SU(2)}(g^L(g^R)^{-1})$ where $(g^L,g^R)\in SU(2)\times SU(2)$. It is invariant
under $SU(2)$ multiplication (of the diagonal subgroup in $SU(2)\times SU(2)$) from the left and
from the right. This is a stronger symmetry than in the general case where $E$ is just required to
be invariant under the adjoint action of the subgroup $\H$.  Because of this enhanced symmetry the
effective face weights just depend on variables  $G\times G/\H$, associated to every half edge.  The
subspace of the Hilbert space spanned by basis states with $k=0$ is exactly the subspace invariant
under this $\H$ group action.

Let us turn to the $W$ operator and see in which sense it is  a constraint  on the curvature (as is the case in BF theory). The effective face weights can be computed to
\ba\label{35}
\omega'_{f}(\kappa_{e_1v_1} \kappa_{v_1e_2}, \kappa_{e_2v_2} \kappa_{v_2e_3}, \ldots)
=
\int \left(\prod_{I=1}^N d\gamma_I  \right) \,\,
\delta_{\H} (   \,\, \prod_{J=1}^N \gamma_J \kappa_{e_Jv_J} \kappa_{v_J e_{J+1}} \gamma_J^{-1}
\,\,) \q .
\ea
Here $\gamma_I$ are group variables in $\H=SU(2)$ and $\kappa_{ve}$ abbreviates $g^L_{ve}
(g^R_{ve})^{-1}$ the product of left and (inverse) right copy of the group. $N$ denotes the number
of edges in the face $f$.

The requirement for a non-zero face weight is that there exists  a set of  group elements $\kappa_I'$, which have to be   in the conjugacy class of  $\kappa_I:=\kappa_{e_Iv_I} \kappa_{v_I e_{I+1}}$, such that the product of the $\kappa_I'$  is equal to the identity. For faces with more than two edges  this condition is generically satisfied and the effective face weights only vanish on measure zero submanifolds in the configuration space. Thus we cannot read off a curvature constraint for the general case.

\subsection{Discussion of transfer operator and its limit}

A question of intense research is the relation between the dynamics as defined by the spin foam models on the one hand and the dynamics as defined by the Hamiltonian constraints in loop quantum gravity \cite{thiemannh} on the other hand \cite{Noui:2004iy,Alesci:2010gb,zipfel,bonzomrec,bonzom4d}. The model example is 3D gravity, which is equivalent to BF theory \cite{Noui:2004iy}. In this case the transfer operator is a projector and its image can be described by quantum constraints which reflect the diffeomorphism symmetry of the model. These constraints can also be encoded into recursion relations (in the spin representation) which can be interpreted as the Wheeler DeWitt equations of 3D quantum gravity \cite{bc3d,bonzom3d}.

In 4D gravitational models the situation is complicated by two main issues. One is that diffeomorphism symmetry in discrete 4D gravity models are generically broken even on the classical level, so that one cannot expect  the transfer operators to be a pure projector and to lead to constraints \cite{b1,bd1,hoehn1}. There is an exception to this general picture, in cases that the dynamics allows only for flat geometries. This is the case for special triangulations described in \cite{ryan1,hoehn2} and more specifically for so called tent moves at four--valent vertices \cite{hoehn1}, on which we will comment more below.

The second main issue is the appearance of simplicity constraints. The natural boundary Hilbert
spaces for the current spin foam models are based on $G=SO(4)$ (or $G=SO(3,1)$) holonomies, whereas
the LQG Hilbert space is based on $\H=SU(2)$ holonomies. Although we can formally define the
transfer operator (i.e. for the EPRL model) as an operator on the LQG Hilbert space it rather
involves the multiplication operator $W$ with $G$ holonomies.

Let us comment more on the two issues and point out avenues for further research.  In case of broken diffeomorphism symmetry the transfer operator is rather an evolution operator corresponding to some finite time step instead of a projection operator. Thus one could ask for the limit in which this finite time step is taken to be small in order to extract the time generator, that is the Hamiltonian. Indeed this is the standard procedure for lattice gauge theory, in which the time step is encoded in the lattice constant (in time direction). In gravity however we do not have such an explicit lattice constant (or other coupling constant). Rather the time distance is boundary state dependent and might only emerge semiclassically. That is to take the limit of infinitesimal time steps we need to entangle this procedure with some semi-classical limit. We discussed different boundary Hilbert spaces, hence different types of semi--classical states are possible, including Hall--like semiclassical states \cite{winklerthiemann, bahrthiemann} adapted to the universal boundary Hilbert space and coherent states used in the discussion of the semi--classical limit of spin foam models \cite{Barrett:2009gg,Barrett:2009mw,Barrett:2009as,Barrett:2009cj,Barrett:2010ex,Hellmann:2010nf,Conrady:2008mk}.

An alternative to considering the limit of infinitesimal small time like distances is to `improve' the transfer operator by basically coarse graining \cite{bd2,Steinhaus,tate}, i.e. considering an effective transfer operator $T'=T^N$. This again is similar to standard statistical lattice theories, where the limit $T^N$ for $N\rightarrow \infty$ leads to a projector on the eigenspace corresponding to the highest eigenvalue of $T$. However, in general some scaling is required in order to obtain an interesting projector, i.e. the highest eigenvalue should be rather highly degenerate. To this end it might be necessary to  refine also the spatial discreteness, as the limit of continuous time but discrete space might rather not lead to a restoration of diffeomorphism symmetry \cite{morse}.

Here we discussed a global transfer operator which acts on the entire hypersurface. An alternative, more adapted to the `multi--fingered' time evolutions are so--called tent moves, discussed in \cite{tent,b1,bd1,hoehn1}, which rather evolve  single vertices in the hypersurface. To define the corresponding transfer operator one needs to provide the possibility of gluing wedges, which is discussed in section \ref{glue glue}. Tent moves are especially interesting as these would lead to localized, vertex based (Hamiltonian) constraints (in case the symmetries are realized). Indeed for classical Regge calculus tent moves at four--valent edges lead  to constraints as the simplicial geometery remains flat in this case \cite{hoehn1,hoehn2} and diffeomorphism symmetry is preserved. Here it would be interesting to know if the spin foam  transfer operator corresponding  to a tent move at a four--valent vertex leads to a projector and therefore constraints or not. More generally the question is whether we can obtain a dynamics describing flat geometries for triangulations that in Regge calculus only allow for flat metrics due to topological reasons \cite{ryan1}.

The second more technical main point concerns the
simplicity constraints. As we have seen the transfer operator naturally involves $W$ as a $G$--group
holonomy and boundary Hilbert spaces based on $G$. Hence, even if we can formally define matrix
elements of the transfer operator on the LQG Hilbert space (based on $\H$ holonomies), a comparison
to a (Hamiltonian) operator expressed in terms of $\H$ holonomies and fluxes is rather difficult. An
alternative is provided by the recent work \cite{bod1,bod2} which provides a canonical connection
formulation based on $G$ holonomies. In this case the Hamiltonian constraints have to be augmented
by a term that makes them gauge invariant with respect to the (primary) simplicity constraints. One
could argue that this term is taken care off in spin foams, as $K_e$ (or $K_e^{1/2}$ or $\mu$)
projects back onto the solutions of the (primary) simplicity constraints. However, heuristically the
transfer operator is the exponential of the Hamiltonians. The additional term in the continuum
Hamiltonian constraints takes care of staying on the simplicity constraint hypersurface at all
times.  In contrast, with the spin foam transfer operator we rather project onto the simplicity
constraint hypersurface in-between discrete time steps. The discrete time steps itself involve $G$
group holonomies which in general map out of the subspace defined by the simplicity constraints.
This is reminiscent of discussions in \cite{alexandrovproj}, which argues that simplicity projectors
should be inserted at each point of the $G$--holonomies. More generally this problem is connected
with the issue of how secondary simplicity constraints \cite{alexandrovr,zapata,ryan1,ryan2} are
implemented into spin foams. The work here offers the possibility to check in which sense the
imposition of primary simplicity constraints at consecutive time steps leads to an imposition of
secondary simplicity constraints, as for instance argued in \cite{etera06}. These insights might
allow a relation between $G$ holonomies and holonomies involving the Ashtekar--Barbero connection
\cite{wieland}, which in turn will ease the comparison of the transfer operator with the Hamiltonian
constraints.

\section{Discussion}\label{discussion}

We have shown that the new holonomy representation presented in this and the companion paper \cite{Comp} provides several advantages. There is a clear parametrization of the space of models. From the parameters one can easily read off the reality of the amplitudes, and it turns out that the BC, EPRL, FK and BO model have real amplitudes in this representation. The holonomy models lead to a natural boundary Hilbert space, which is introduced in this work. This universal boundary Hilbert space is the same Hilbert space for different choices of simplicity functions $E$. Thus different possibilities to impose the simplicity constraints can be compared in one and the same Hilbert space. Furthermore, as the models can be naturally defined on arbitrary  2--complexes we also can obtain arbitrary (and not only 4--valent) graphs on which the boundary Hilbert spaces are based. A (dynamical) notion of cylindrical consistency, related to coarse graining,  can also be introduced on these Hilbert spaces \cite{Bahr:2011aa,bd1205}.

The universal boundary Hilbert space results from a definition of `equal time' slices which is most natural for lattice gauge theory. Alternatively we can adopt slices and gluings more custom to spinfoams and obtain as boundary Hilbert space the Hilbert space of projected spin networks. We detailed the conditions under which such map between the different slicings is possible and constructed the corresponding $\mu$--map. These considerations allowed also a discussion on the different kind of basic building blocks, i.e. faces, half faces and wedges, and the possible gluings which allow a combination of these building blocks to the full partition function.

We explicitly constructed the representation of different current spin foam models in our new representation -- the main difference between the models are the simplicity functions $E$, that imposes the simplicity constraints.

Finally, we derived a general form of the transfer operator on the different boundary Hilbert
spaces. On the universal Hilbert space this form is given by $T=K_0^{1/2}P_GWP_GK_0^{1/2}$ with $W$
representing  a product over holonomy operators over faces, $P_G$ is the projector on the $G$
invariant subspace and $K_0^{1/2}$ can be seen as imposing the simplicity constraints on the Hilbert
space. Similarly on the Hilbert space of projected spin network functions we have $T'=\mu^\dagger
W\mu$ with $\mu,\mu^\dagger$ taking over the role of $K$ to impose the simplicity constraints. This
form might shed some light on the discussion of how to best impose simplicity constraints into spin
foams \cite{alexandrovr,ryan2,bod2,Geiller:2011aa}. In the current models the secondary simplicity
constraints are not imposed. It is argued that the imposition of primary simplicity constraints on
each time step should also lead to the imposition of secondary constraints. (Note however that even
the primary constraints are imposed weakly in the EPRL type models.) Indeed, here this is made
obvious in the form of the transfer operator, who is projected by either $K_0$ or $\mu$. Of course
the projections are only inserted at discrete time steps and not continuously in time. Related to
this issue is that fundamentally the transfer operator still includes a $G=SO(4)$--holonomy operator
and not a holonomy operator based on the Ashtekar--Barbero connection. A question for future
research is how this form relates to a discretion which starts from the beginning with the
$SU(2)$--Ashtekar Barbero connection, which arises by classicaly solving for the simplicity
constraints \cite{Geiller:2011bh}. This question can be also studied in a semi-classical limit using
semi-classical states in either of the boundary Hilbert spaces or the techniques presented in
\cite{AD}.

\vspace{1cm}

\section*{Acknowledgements}

The authors appreciate intensive discussions with Benjamin Bahr,
BD thanks furthermore Valentin Bonzom for discussions. WK thanks Daniele Oriti for the explanation of the BO model. Research at Perimeter Institute is supported by the Government of Canada through Industry Canada and by the Province of Ontario through the Ministry of Research and Innovation.
\footnotesize{
\bibliography{FSM}
\bibliographystyle{hplain}
}
\end{document}

%% file: Macros.tex
\usepackage{amssymb,amstext,amsmath,amsthm,amsxtra}

\newcommand{\bea}{\begin{eqnarray}}
\newcommand{\eea}{\end{eqnarray}}
\newcommand{\ba}{\begin{eqnarray}}
\newcommand{\ea}{\end{eqnarray}}
\newcommand{\nn}{\nonumber}
\newcommand{\be}{\begin{equation}}
\newcommand{\ee}{\end{equation}}

\theoremstyle{definition}
\newtheorem{defi}{Definition}[section]
\theoremstyle{plain}

\newcommand\q{\quad}

\newcommand{\nb}{{\bf n}}

\newcommand{\HH}{\mathcal{H}}

\newcommand{\R}{\mathbb{R}}
\newcommand{\CC}{\mathcal{C}}
\newcommand{\EE}{\mathcal{E}}

\newcommand{\id}{\mathbf{1}}
\DeclareMathOperator{\tr}{tr}

\newcommand{\ZZ}{\mathcal{Z}}

\newcommand{\Inv}{\rm Inv}

\newcommand{\tensor}{\otimes}
\newcommand{\tens}{\otimes}

\newcommand{\SU}{\mathrm{SU}}

\newcommand{\Spin}{\mathrm{Spin}}
\newcommand{\Hom}{\mathrm{Hom}}

\newcommand{\dd}{\mathrm{d}}

\def\la{\langle}
\def\ra{\rangle}

%% file: HolSFM-Formal.bbl
\begin{thebibliography}{100}

\bibitem{Alesci:2010gb}
Emanuele Alesci and Carlo Rovelli.
\newblock {A Regularization of the hamiltonian constraint compatible with the
  spinfoam dynamics}.
\newblock {\em Phys.Rev.}, D82:044007, 2010, 1005.0817.

\bibitem{zipfel}
Emanuele Alesci, Thomas Thiemann, and Antonia Zipfel.
\newblock {Linking covariant and canonical LQG: New solutions to the Euclidean
  Scalar Constraint}.
\newblock {\em Phys.Rev.}, D86:024017, 2012, 1109.1290.

\bibitem{Alexandrov:2002xc}
Sergei Alexandrov.
\newblock {Hilbert space structure of covariant loop quantum gravity}.
\newblock {\em Phys.Rev.}, D66:024028, 2002, gr-qc/0201087.

\bibitem{alexandrovproj}
Sergei Alexandrov.
\newblock {The new vertices and canonical quantization}.
\newblock {\em Phys.Rev.}, D82:024024, 2010, 1004.2260.

\bibitem{alexandrovr}
Sergei Alexandrov, Marc Geiller, and Karim Noui.
\newblock {Spin Foams and Canonical Quantization}.
\newblock {\em SIGMA}, 8:055, 2012, 1112.1961.

\bibitem{Ashtekar:1991kc}
Abhay Ashtekar and C.J. Isham.
\newblock {Representations of the holonomy algebras of gravity and nonAbelian
  gauge theories}.
\newblock {\em Class.Quant.Grav.}, 9:1433--1468, 1992, hep-th/9202053.

\bibitem{Ashtekar:1994wa}
Abhay Ashtekar and Jerzy Lewandowski.
\newblock {Differential geometry on the space of connections via graphs and
  projective limits}.
\newblock {\em J.Geom.Phys.}, 17:191--230, 1995, hep-th/9412073.

\bibitem{Ashtekar:1994mh}
Abhay Ashtekar and Jerzy Lewandowski.
\newblock {Projective techniques and functional integration for gauge
  theories}.
\newblock {\em J.Math.Phys.}, 36:2170--2191, 1995, gr-qc/9411046.

\bibitem{Ashtekar:1995zh}
Abhay Ashtekar, Jerzy Lewandowski, Donald Marolf, Jose Mourao, and Thomas
  Thiemann.
\newblock {Quantization of diffeomorphism invariant theories of connections
  with local degrees of freedom}.
\newblock {\em J.Math.Phys.}, 36:6456--6493, 1995, gr-qc/9504018.

\bibitem{Baez:1997zt}
John~C. Baez.
\newblock {Spin foam models}.
\newblock {\em Class. Quant. Grav.}, 15:1827--1858, 1998, gr-qc/9709052.

\bibitem{Baez:1999sr}
John~C. Baez.
\newblock {An introduction to spin foam models of BF theory and quantum
  gravity}.
\newblock {\em Lect. Notes Phys.}, 543:25--94, 2000, gr-qc/9905087.

\bibitem{Bahr:2011aa}
Benjamin Bahr.
\newblock {Operator Spin Foams: holonomy formulation and coarse graining}.
\newblock 2011, 1112.3567.

\bibitem{bd1}
Benjamin Bahr and Bianca Dittrich.
\newblock {(Broken) Gauge Symmetries and Constraints in Regge Calculus}.
\newblock {\em Class.Quant.Grav.}, 26:225011, 2009, 0905.1670.

\bibitem{bd2}
Benjamin Bahr and Bianca Dittrich.
\newblock {Improved and Perfect Actions in Discrete Gravity}.
\newblock {\em Phys.Rev.}, D80:124030, 2009, 0907.4323.

\bibitem{Comp}
Benjamin Bahr, Bianca Dittrich, Frank Hellmann, and Wojciech Kaminski.
\newblock {Holonomy Spin Foam Models: Definition and Coarse Graining}.
\newblock 2012, 1208.3388.

\bibitem{Bahr:2011yc}
Benjamin Bahr, Bianca Dittrich, and James~P. Ryan.
\newblock {Spin foam models with finite groups}.
\newblock 2011, 1103.6264.

\bibitem{Steinhaus}
Benjamin Bahr, Bianca Dittrich, and Sebastian Steinhaus.
\newblock {Perfect discretization of reparametrization invariant path
  integrals}.
\newblock {\em Phys.Rev.}, D83:105026, 2011, 1101.4775.

\bibitem{Bahr:2010bs}
Benjamin Bahr, Frank Hellmann, Wojciech Kaminski, Marcin Kisielowski, and Jerzy
  Lewandowski.
\newblock {Operator Spin Foam Models}.
\newblock {\em Class.Quant.Grav.}, 28:105003, 2011, 1010.4787.

\bibitem{bahrthiemann}
Benjamin Bahr and Thomas Thiemann.
\newblock {Gauge-invariant coherent states for loop quantum gravity. II.
  Non-Abelian gauge groups}.
\newblock {\em Class.Quant.Grav.}, 26:045012, 2009, 0709.4636.

\bibitem{Baratin:2011tx}
Aristide Baratin and Daniele Oriti.
\newblock {Quantum simplicial geometry in the group field theory formalism:
  reconsidering the Barrett-Crane model}.
\newblock {\em New J.Phys.}, 13:125011, 2011, 1108.1178.

\bibitem{Baratin:2011hp}
Aristide Baratin and Daniele Oriti.
\newblock {Group field theory and simplicial gravity path integrals: A model
  for Holst-Plebanski gravity}.
\newblock {\em Phys.Rev.}, D85:044003, 2012, 1111.5842.

\bibitem{bc3d}
John~W. Barrett and Louis Crane.
\newblock {An Algebraic interpretation of the Wheeler-DeWitt equation}.
\newblock {\em Class.Quant.Grav.}, 14:2113--2121, 1997, gr-qc/9609030.

\bibitem{Barrett:1997gw}
John~W. Barrett and Louis Crane.
\newblock {Relativistic spin networks and quantum gravity}.
\newblock {\em J.Math.Phys.}, 39:3296--3302, 1998, gr-qc/9709028.

\bibitem{Barrett:1999qw}
John~W. Barrett and Louis Crane.
\newblock {A Lorentzian signature model for quantum general relativity}.
\newblock {\em Class.Quant.Grav.}, 17:3101--3118, 2000, gr-qc/9904025.

\bibitem{Barrett:2009cj}
John~W. Barrett, R.J. Dowdall, Winston~J. Fairbairn, Henrique Gomes, and Frank
  Hellmann.
\newblock {A Summary of the asymptotic analysis for the EPRL amplitude}.
\newblock 2009, 0909.1882.

\bibitem{Barrett:2009gg}
John~W. Barrett, R.J. Dowdall, Winston~J. Fairbairn, Henrique Gomes, and Frank
  Hellmann.
\newblock {Asymptotic analysis of the EPRL four-simplex amplitude}.
\newblock {\em J.Math.Phys.}, 50:112504, 2009, 0902.1170.

\bibitem{Barrett:2010ex}
John~W. Barrett, R.J. Dowdall, Winston~J. Fairbairn, Henrique Gomes, Frank
  Hellmann, et~al.
\newblock {Asymptotics of 4d spin foam models}.
\newblock {\em Gen.Rel.Grav.}, 43:2421--2436, 2011, 1003.1886.

\bibitem{Barrett:2009mw}
John~W. Barrett, R.J. Dowdall, Winston~J. Fairbairn, Frank Hellmann, and
  Roberto Pereira.
\newblock {Lorentzian spin foam amplitudes: Graphical calculus and
  asymptotics}.
\newblock {\em Class.Quant.Grav.}, 27:165009, 2010, 0907.2440.

\bibitem{Barrett:2009as}
John~W. Barrett, Winston~J. Fairbairn, and Frank Hellmann.
\newblock {Quantum gravity asymptotics from the SU(2) 15j symbol}.
\newblock {\em Int.J.Mod.Phys.}, A25:2897--2916, 2010, 0912.4907.

\bibitem{Barrett:2011qe}
John~W. Barrett and Frank Hellmann.
\newblock {Holonomy observables in Ponzano-Regge type state sum models}.
\newblock {\em Class.Quant.Grav.}, 29:045006, 2012, 1106.6016.

\bibitem{Geloun:2010vj}
Joseph Ben~Geloun, Razvan Gurau, and Vincent Rivasseau.
\newblock {EPRL/FK Group Field Theory}.
\newblock {\em Europhys.Lett.}, 92:60008, 2010, 1008.0354.

\bibitem{Bianchi:2012nk}
Eugenio Bianchi and Frank Hellmann.
\newblock {The Construction of Spin Foam Vertex Amplitudes}.
\newblock 2012, 1207.4596.

\bibitem{bod1}
Norbert Bodendorfer, Thomas Thiemann, and Andreas Thurn.
\newblock {New Variables for Classical and Quantum Gravity in all Dimensions
  III. Quantum Theory}.
\newblock 2011, 1105.3705.

\bibitem{bod2}
Norbert Bodendorfer, Thomas Thiemann, and Andreas Thurn.
\newblock {On the Implementation of the Canonical Quantum Simplicity
  Constraint}.
\newblock 2011, 1105.3708.

\bibitem{Bonzom:2009wm}
Valentin Bonzom.
\newblock {From lattice BF gauge theory to area-angle Regge calculus}.
\newblock {\em Class.Quant.Grav.}, 26:155020, 2009, 0903.0267.

\bibitem{Bonzom:2009hw}
Valentin Bonzom.
\newblock {Spin foam models for quantum gravity from lattice path integrals}.
\newblock {\em Phys.Rev.}, D80:064028, 2009, 0905.1501.

\bibitem{bonzom4d}
Valentin Bonzom.
\newblock {Spin foam models and the Wheeler-DeWitt equation for the quantum
  4-simplex}.
\newblock {\em Phys.Rev.}, D84:024009, 2011, 1101.1615.

\bibitem{bonzom3d}
Valentin Bonzom and Laurent Freidel.
\newblock {The Hamiltonian constraint in 3d Riemannian loop quantum gravity}.
\newblock {\em Class.Quant.Grav.}, 28:195006, 2011, 1101.3524.

\bibitem{bonzomrec}
Valentin Bonzom, Etera~R. Livine, and Simone Speziale.
\newblock {Recurrence relations for spin foam vertices}.
\newblock {\em Class.Quant.Grav.}, 27:125002, 2010, 0911.2204.

\bibitem{Boulatov:1992vp}
D.V. Boulatov.
\newblock {A Model of three-dimensional lattice gravity}.
\newblock {\em Mod.Phys.Lett.}, A7:1629--1646, 1992, hep-th/9202074.

\bibitem{Conrady:2008mk}
Florian Conrady and Laurent Freidel.
\newblock {On the semiclassical limit of 4d spin foam models}.
\newblock {\em Phys.Rev.}, D78:104023, 2008, 0809.2280.

\bibitem{DePietri:1999bx}
Roberto De~Pietri, Laurent Freidel, Kirill Krasnov, and Carlo Rovelli.
\newblock {Barrett-Crane model from a Boulatov-Ooguri field theory over a
  homogeneous space}.
\newblock {\em Nucl.Phys.}, B574:785--806, 2000, hep-th/9907154.

\bibitem{tate}
B.~Dittrich and K.~Tate.
\newblock {Exact discretizations by truncating the space of solutions}.
\newblock {\em in preparation.}

\bibitem{b1}
Bianca Dittrich.
\newblock {Diffeomorphism symmetry in quantum gravity models}.
\newblock {\em Adv.Sci.Lett.}, 2:151--164, 2009, 0810.3594.

\bibitem{bd1205}
Bianca Dittrich.
\newblock {From the discrete to the continuous: Towards a cylindrically
  consistent dynamics}.
\newblock 2012, 1205.6127.

\bibitem{bd12}
Bianca Dittrich.
\newblock {How to construct diffeomorphism symmetry on the lattice}.
\newblock 2012, 1201.3840.

\bibitem{Dittrich:2011zh}
Bianca Dittrich, Frank~C. Eckert, and Mercedes Martin-Benito.
\newblock {Coarse graining methods for spin net and spin foam models}.
\newblock {\em New J.Phys.}, 14:035008, 2012, 1109.4927.

\bibitem{hoehn1}
Bianca Dittrich and Philipp~A. Hohn.
\newblock {From covariant to canonical formulations of discrete gravity}.
\newblock {\em Class.Quant.Grav.}, 27:155001, 2010, 0912.1817.

\bibitem{hoehn2}
Bianca Dittrich and Philipp~A. Hohn.
\newblock {Canonical simplicial gravity}.
\newblock {\em Class.Quant.Grav.}, 29:115009, 2012, 1108.1974.

\bibitem{ryan2}
Bianca Dittrich and James~P. Ryan.
\newblock {Simplicity in simplicial phase space}.
\newblock {\em Phys.Rev.}, D82:064026, 2010, 1006.4295.

\bibitem{ryan1}
Bianca Dittrich and James~P. Ryan.
\newblock {Phase space descriptions for simplicial 4d geometries}.
\newblock {\em Class.Quant.Grav.}, 28:065006, 2011, 0807.2806.

\bibitem{Dupuis:2010jn}
Maite Dupuis and Etera~R. Livine.
\newblock {Lifting SU(2) Spin Networks to Projected Spin Networks}.
\newblock {\em Phys.Rev.}, D82:064044, 2010, 1008.4093.

\bibitem{Engle:2007wy}
Jonathan Engle, Etera Livine, Roberto Pereira, and Carlo Rovelli.
\newblock {LQG vertex with finite Immirzi parameter}.
\newblock {\em Nucl. Phys.}, B799:136--149, 2008, 0711.0146.

\bibitem{Engle:2007uq}
Jonathan Engle, Roberto Pereira, and Carlo Rovelli.
\newblock {The loop-quantum-gravity vertex-amplitude}.
\newblock {\em Phys. Rev. Lett.}, 99:161301, 2007, 0705.2388.

\bibitem{Engle:2007qf}
Jonathan Engle, Roberto Pereira, and Carlo Rovelli.
\newblock {Flipped spinfoam vertex and loop gravity}.
\newblock {\em Nucl. Phys.}, B798:251--290, 2008, 0708.1236.

\bibitem{Freidel:1998pt}
Laurent Freidel and Kirill Krasnov.
\newblock {Spin foam models and the classical action principle}.
\newblock {\em Adv.Theor.Math.Phys.}, 2:1183--1247, 1999, hep-th/9807092.

\bibitem{Freidel:2007py}
Laurent Freidel and Kirill Krasnov.
\newblock {A New Spin Foam Model for 4d Gravity}.
\newblock {\em Class. Quant. Grav.}, 25:125018, 2008, 0708.1595.

\bibitem{Gambini:2005sv}
Rodolfo Gambini and Jorge Pullin.
\newblock {Classical and quantum general relativity: A New paradigm}.
\newblock {\em Gen.Rel.Grav.}, 37:1689--1694, 2005, gr-qc/0505052.

\bibitem{Gambini:2005vn}
Rodolfo Gambini and Jorge Pullin.
\newblock {Consistent discretization and canonical classical and quantum Regge
  calculus}.
\newblock {\em Int.J.Mod.Phys.}, D15:1699--1706, 2006, gr-qc/0511096.

\bibitem{Geiller:2011bh}
Marc Geiller, Marc Lachieze-Rey, and Karim Noui.
\newblock {A new look at Lorentz-Covariant Loop Quantum Gravity}.
\newblock {\em Phys.Rev.}, D84:044002, 2011, 1105.4194.

\bibitem{Geiller:2011aa}
Marc Geiller and Karim Noui.
\newblock {Testing the imposition of the Spin Foam Simplicity Constraints}.
\newblock {\em Class.Quant.Grav.}, 29:135008, 2012, 1112.1965.

\bibitem{GT}
I.M. Gelfand and M.L. Tsetlin.
\newblock Finite-dimensional representations of the group of unimodular
  matrices.
\newblock {\em Dokl. Akad. Nauk SSSR}, 71:825–828, 1950.
\newblock English transl. I.M. Gelfand, Collected Papers, Vol. II,
  Springer-Verlag (1988), 653–656.

\bibitem{Gurau:2009tw}
Razvan Gurau.
\newblock {Colored Group Field Theory}.
\newblock {\em Commun.Math.Phys.}, 304:69--93, 2011, 0907.2582.

\bibitem{Hellmann:2010nf}
Frank Hellmann.
\newblock {State Sums and Geometry}.
\newblock 2010, 1102.1688.

\bibitem{AD}
Frank Hellmann and Wojciech Kaminski.
\newblock Asymptotic dynamics: Large scale behaviour of spin foam partition
  functions.

\bibitem{Kaminski:2009fm}
Wojciech Kaminski, Marcin Kisielowski, and Jerzy Lewandowski.
\newblock {Spin-Foams for All Loop Quantum Gravity}.
\newblock {\em Class.Quant.Grav.}, 27:095006, 2010, 0909.0939.

\bibitem{Kaminski:2009cc}
Wojciech Kaminski, Marcin Kisielowski, and Jerzy Lewandowski.
\newblock {The EPRL intertwiners and corrected partition function}.
\newblock {\em Class.Quant.Grav.}, 27:165020, 2010, 0912.0540.

\bibitem{kauffman1994temperley}
L.H. Kauffman and S.~Lins.
\newblock {\em Temperley-Lieb Recoupling Theory and Invariants of 3-Manifolds
  (AM-134)}.
\newblock Annals of Mathematics Studies. Princeton University Press, 1994.

\bibitem{ks}
John Kogut and Leonard Susskind.
\newblock Hamiltonian formulation of wilson's lattice gauge theories.
\newblock {\em Phys. Rev. D}, 11:395--408, Jan 1975.

\bibitem{kr}
John~B. Kogut.
\newblock An introduction to lattice gauge theory and spin systems.
\newblock {\em Rev. Mod. Phys.}, 51:659--713, Oct 1979.

\bibitem{Lewandowski:2005jk}
Jerzy Lewandowski, Andrzej Okolow, Hanno Sahlmann, and Thomas Thiemann.
\newblock {Uniqueness of diffeomorphism invariant states on holonomy-flux
  algebras}.
\newblock {\em Commun.Math.Phys.}, 267:703--733, 2006, gr-qc/0504147.

\bibitem{Livine:2002ak}
Etera~R Livine.
\newblock {Projected spin networks for Lorentz connection: Linking spin foams
  and loop gravity}.
\newblock {\em Class.Quant.Grav.}, 19:5525--5542, 2002, gr-qc/0207084.

\bibitem{etera06}
Etera~R. Livine.
\newblock {Towards a Covariant Loop Quantum Gravity}.
\newblock 2006, gr-qc/0608135.

\bibitem{Livine:2002rh}
Etera~R. Livine and Daniele Oriti.
\newblock {Implementing causality in the spin foam quantum geometry}.
\newblock {\em Nucl.Phys.}, B663:231--279, 2003, gr-qc/0210064.

\bibitem{morse}
P.A.. Morse.
\newblock {Approximate diffeomorphism invariance in near flat simplicial
  geometries}.
\newblock {\em Class.Quant.Grav.}, 9:2489, 1992.

\bibitem{Noui:2004ja}
Karim Noui and Alejandro Perez.
\newblock {Dynamics of loop quantum gravity and spin foam models in
  three-dimensions}.
\newblock pages 648--654, 2004, gr-qc/0402112.

\bibitem{Noui:2004iy}
Karim Noui and Alejandro Perez.
\newblock {Three-dimensional loop quantum gravity: Physical scalar product and
  spin foam models}.
\newblock {\em Class.Quant.Grav.}, 22:1739--1762, 2005, gr-qc/0402110.

\bibitem{Oeckl:2005rh}
R.~Oeckl.
\newblock {Discrete gauge theory: From lattices to TQFT}.
\newblock London, UK: Imperial College Pr. (2005) 202 p.

\bibitem{Oeckl:2001wm}
Robert Oeckl.
\newblock {Generalized lattice gauge theory, spin foams and state sum
  invariants}.
\newblock {\em J.Geom.Phys.}, 46:308--354, 2003, hep-th/0110259.

\bibitem{Ooguri:1992eb}
Hirosi Ooguri.
\newblock {Topological lattice models in four-dimensions}.
\newblock {\em Mod.Phys.Lett.}, A7:2799--2810, 1992, hep-th/9205090.

\bibitem{Oriti:2002hv}
Daniele Oriti.
\newblock {Boundary terms in the Barrett-Crane spin foam model and consistent
  gluing}.
\newblock {\em Phys.Lett.}, B532:363--372, 2002, gr-qc/0201077.

\bibitem{Oriti:2006se}
Daniele Oriti.
\newblock {The Group field theory approach to quantum gravity}.
\newblock 2006, gr-qc/0607032.

\bibitem{Oriti:2000hh}
Daniele Oriti and Ruth~M. Williams.
\newblock {Gluing 4 simplices: A Derivation of the Barrett-Crane spin foam
  model for Euclidean quantum gravity}.
\newblock {\em Phys.Rev.}, D63:024022, 2001, gr-qc/0010031.

\bibitem{Perez:2012wv}
Alejandro Perez.
\newblock {The Spin Foam Approach to Quantum Gravity}.
\newblock 2012, 1205.2019.

\bibitem{Perez:2000fs}
Alejandro Perez and Carlo Rovelli.
\newblock {A Spin foam model without bubble divergences}.
\newblock {\em Nucl.Phys.}, B599:255--282, 2001, gr-qc/0006107.

\bibitem{Perez:2000ec}
Alejandro Perez and Carlo Rovelli.
\newblock {Spin foam model for Lorentzian general relativity}.
\newblock {\em Phys.Rev.}, D63:041501, 2001, gr-qc/0009021.

\bibitem{Pfeiffer:2001ny}
Hendryk Pfeiffer.
\newblock {Dual variables and a connection picture for the Euclidean
  Barrett-Crane model}.
\newblock {\em Class.Quant.Grav.}, 19:1109--1138, 2002, gr-qc/0112002.

\bibitem{Reisenberger:2000fy}
Michael Reisenberger and Carlo Rovelli.
\newblock {Spin foams as Feynman diagrams}.
\newblock pages 431--448, 2000, gr-qc/0002083.

\bibitem{Reisenberger:1994aw}
Michael~P. Reisenberger.
\newblock {World sheet formulations of gauge theories and gravity}.
\newblock 1994, gr-qc/9412035.

\bibitem{Reisenberger:1997sk}
Michael~P. Reisenberger.
\newblock {A Lattice world sheet sum for 4-d Euclidean general relativity}.
\newblock 1997, gr-qc/9711052.

\bibitem{Reisenberger:1996pu}
Michael~P Reisenberger and Carlo Rovelli.
\newblock {'Sum over surfaces' form of loop quantum gravity}.
\newblock {\em Phys.Rev.}, D56:3490--3508, 1997, gr-qc/9612035.

\bibitem{Reisenberger:2000zc}
Michael~P. Reisenberger and Carlo Rovelli.
\newblock {Space-time as a Feynman diagram: The Connection formulation}.
\newblock {\em Class.Quant.Grav.}, 18:121--140, 2001, gr-qc/0002095.

\bibitem{Rovelli:1989za}
Carlo Rovelli and Lee Smolin.
\newblock {Loop Space Representation of Quantum General Relativity}.
\newblock {\em Nucl.Phys.}, B331:80, 1990.

\bibitem{Rovelli:1995ac}
Carlo Rovelli and Lee Smolin.
\newblock {Spin networks and quantum gravity}.
\newblock {\em Phys.Rev.}, D52:5743--5759, 1995, gr-qc/9505006.

\bibitem{smit}
J.~Smit.
\newblock {\em Introduction to Quantum Fields on a Lattice}.
\newblock Cambridge Lecture Notes in Physics. Cambridge University Press, 2002.

\bibitem{tent}
R.~Sorkin.
\newblock {Time Evolution Problem In Regge Calculus}.
\newblock {\em Phys.Rev.}, D12:385, 1975.

\bibitem{wieland}
Simone Speziale and Wolfgang~M. Wieland.
\newblock {The twistorial structure of loop-gravity transition amplitudes}.
\newblock 2012, 1207.6348.

\bibitem{thiemannh}
T.~Thiemann.
\newblock {Quantum spin dynamics (QSD)}.
\newblock {\em Class.Quant.Grav.}, 15:839--873, 1998, gr-qc/9606089.

\bibitem{winklerthiemann}
T.~Thiemann and O.~Winkler.
\newblock {Gauge field theory coherent states (GCS). 2. Peakedness properties}.
\newblock {\em Class.Quant.Grav.}, 18:2561--2636, 2001, hep-th/0005237.

\bibitem{zapata}
Jose~A. Zapata.
\newblock {Topological lattice gravity using selfdual variables}.
\newblock {\em Class.Quant.Grav.}, 13:2617--2634, 1996, gr-qc/9603030.

\end{thebibliography}
